\documentclass[%
reprint,
superscriptaddress,
nofootinbib,
twocolumn,
notitlepage,
amsmath,amssymb,
aps,
pra
]{revtex4-2}

\usepackage{graphicx}
\usepackage{dcolumn}
\usepackage{bm}
\usepackage{tablefootnote}

\usepackage{amsthm}
\usepackage{mathtools}
\usepackage{amsmath}
\usepackage{amssymb}
\usepackage{graphicx}
\usepackage{url}
\usepackage{float}
\usepackage{bm}
\usepackage{ifthen}
\usepackage[usenames,dvipsnames]{color}
\usepackage{mathrsfs}
\usepackage[colorlinks=true,citecolor=blue,urlcolor=black]{hyperref}
\usepackage{float}
\usepackage{subfigure}
\usepackage{enumitem}
\usepackage{color}
\usepackage{setspace}
\usepackage{braket}
\usepackage{comment}
\usepackage{lipsum}

\newcommand{\be}{\begin{equation}}
	\newcommand{\ee}{\end{equation}}

\newcommand{\blfootnote}[1]{
	\begingroup
	\renewcommand{\thefootnote}{}
	\footnotetext{#1}
	\addtocounter{footnote}{-1}
	\endgroup	
}

\DeclareMathOperator{\Perm}{Perm}

\begin{document}
	
	\title{Decoy-state optical quantum information processing with coherent states}
	
		\author{Wenyuan Wang}
	\thanks{The authors contributed equally to this work.}
	\affiliation{Information Science Division, Nara Institute of Science and Technology, Ikoma, Nara, Japan}
	\affiliation{Department of Physics and Astronomy, University of Calgary, Calgary, Canada}
		
	\author{H. F. Chau}
	\thanks{The authors contributed equally to this work.}
	\affiliation{Department of Physics, University of Hong Kong, Pokfulam Road, Hong Kong}

	\begin{abstract}
		
		Photonic qubits play an important role in quantum information processing. However, to date, efficient high-speed single photon sources and photon number resolving (PNR) detectors are still difficult to make. Here, we use phase-randomized coherent states and threshold detectors, combined with decoy-state post-processing, to simulate single photon sources and PNR detectors and perform various quantum information processing tasks. We propose a simple linear regression method to predict Fock state statistics based on observable statistics from coherent-state probes, combined with random/structured sampling of probe settings. Compared with traditional decoy-state post-processing based on linear programming and Cartesian product grid sampling, our method enables drastically improved accuracy and much reduced probe resources. It works particularly well for unknown unitary circuits of interest with fixed-size Hilbert space dimensions. For instance, it can accurately simulate 4-photon interference in linear optics with just 32 input/output probe settings and 6-photon interference with 256 probe settings. Moreover, it also works for non-unitary (such as lossy, nonlinear, or arbitrary) quantum channels. This enables a wide variety of applications that can be implemented with easily attainable coherent light sources and threshold detectors, such as characterization of small-scale linear optical quantum computing circuits or quantum sensing/metrology devices.
		
	\end{abstract}
	\date{\today}
	\maketitle
	\blfootnote{Emails: wenyuan.wang@ucalgary.ca; hfchau@hku.hk.}
	
	\section{Introduction}
	
	Photons are both ideal carriers for long-distance quantum communications and also easy-to-manipulate media for quantum computing. Photonic quantum information processing has seen much development in the past 25 years, from the proposal of linear optical quantum computing (LOQC) \cite{KLM}, to measurement-based quantum computing \cite{onewaycomputing}, to the more recent demonstrations of quantum supremacy with boson sampling \cite{bosonsampling1,bosonsampling2}. Many of those applications require the preparation of single photons or entangled photonic states. Foundational quantum tasks such as Bell test \cite{bell1,bell2,bell3} or quantum applications like metrology \cite{metrology} also require single photons, EPR pairs, or even NOON states. However, it is rather difficult to make an ideal single photon source. Traditional coherent light has a large portion of multi-photon components. Spontaneous parametric down conversion (SPDC) sources are commonly used as heralded single photon sources, but they contain multiphotons too and are also rather inefficient and slow (and emit photons at non-fixed intervals). In recent years, novel single photon sources like quantum dots or even diamond center sources have seen increasing exposure \cite{quantumdot,diamondcenter}, but they are still some distance from being mature and available off-the-shelf (with challenges such as in clock rate, difficulty of maintenance, and being not available at telecom wavelength to extend to quantum communication applications).

	In 2016 and 2018, respectively, two seminal papers \cite{Xiongfeng_interference,Marcos_interference} proposed an innovative approach to borrow the decoy state technique \cite{decoystate1,decoystate2,decoystate3} from the field of quantum cryptography and apply it to optical quantum information processing to simulate single (or higher-order Fock state) photon inputs using only coherent lights\footnote{An unpublished report \cite{Feihu_bosonsampling} in 2014 also proposed to perform optical computing, specifically boson sampling, with decoy states, although the report only considered using heralded SPDC sources and was not focused on coherent lights.}. The decoy state method uses phase-randomized coherent states attenuated to different intensity levels, combined with numerical post-processing, to bound the statistics of fixed-photon-number Fock states entering a circuit.
	
	However, the limitation of Ref. \cite{Xiongfeng_interference} was that it only studied input states (in fact only the specific case with single photons on each input mode), while photon counting detectors are still needed on output ports for full optical quantum computing or information processing.
	
	Ref. \cite{Marcos_interference} addressed this problem by applying decoy state method to detectors too, such that one can use threshold detectors with different levels of attenuation in front of them to simulate photon counting detectors. It also extended the method to the cases of multi-photon inputs/outputs beyond single photons. The main challenge in Ref. \cite{Marcos_interference}, though, was that it used a numerical linear programming (LP) method to solve the Fock state statistics, which suffered from numerical imprecision as the photon number increases in the input/output states to be solved. This severely limited the highest photon numbers that could be calculated. For instance, the largest numerically computed scenario was a tritter with 3 photon inputs, with non-negligible numerical noise.
	
	In this work, we propose to extend the decoy state method into a general framework for optical quantum information processing with coherent states – for not just interferometers, but LOQC gates, boson sampling circuits, metrology setups, and state preparation for quantum communication. These applications naturally require larger circuits and high precision. For this, we propose a simple linear regression method to overcome the main challenge of accurately estimating Fock state statistics from coherent-state observables, enabling us to estimate statistics from larger unitary circuits, such as 4-photon or even up to 6-photon inputs.

	Here we make several important physical observations that make applying the decoy-state method to quantum information processing different from using it in QKD, which are where the effectiveness of our new approach in this paper comes from:

	\begin{enumerate}
		\item \textit{Unitary Circuit Model}: For many user scenarios, such as LOQC, boson sampling, or metrology setups that probe an unknown parameter (such as phase shift) in a circuit, the channel itself is not an arbitrary channel with worst-case adversaries. Despite that the circuit in question is, of course, unknown (otherwise we can just compute all its characteristics on paper), in many cases we often at least know that it is a unitary circuit based on passive linear components, with a known Hilbert space dimension (N input modes and N output modes). This gives us a \text{physical prior} that locks higher-order photon number terms to lower-order terms, since fundamentally they are all determined by the degrees-of-freedom in the unitary circuit itself. Such a physical prior knowledge can greatly enhance the estimation accuracy/stability of high photon number terms.
		
		Note that, the linear regression method we report in this paper can be additionally applied to nonlinear circuits or even arbitrary channels, too, as we will later show in the results. Nonetheless, the method enjoys the highest precision and scalability in scenarios where it can assume a unitary physical prior for the channel.

		\item \textit{Prediction versus Bounding}: For many user applications, we care more about the sampling of output patterns (such as the visibility of interference dips) and their averages rather than an information-theoretically secure upper/lower bound where we assume all possible channel disturbances/attacks. The latter pessimistic assumption is necessary in an adversarial scenario, but it can lead to loosened results in a tabletop setup in one's lab. Here our linear regression method focuses on the accurate \textit{prediction} of Fock state statistics (it also avoids being a machine learning black-box as the predictions are just linear combinations of experimental observables with explicit coefficients). The prediction results, as we will show, are highly accurate and stable, with absolute error values on the order of $10^{-5}$ (with target Fock state statistics outputs in $[0,1]$).
		
		There are, nonetheless, applications where upper/lower bounds are of interest, such as in QKD or state preparation for quantum communication. We further show in our companion paper \cite{this_companion} that a set of provably secure \textit{certified bounds} can be applied to the predictions generated from machine learning, showing that our framework can be extended to quantum cryptography/communication applications. In this framework paper, however, we will focus on the prediction process itself.
		
		\item \textit{Sampling of Probe Settings}: Lastly, for multi-user QKD, such as measurement-device-independent QKD \cite{MDIQKD} or conference key agreement (CKA) \cite{CKA}, the users are not allowed to have pre-shared basis or decoy-state settings, so they are generally forced to each independently select from a list of laser intensities as probes for decoy state analysis. These settings would globally form a Cartesian product grid. However, this sampling method is an \textit{overkill} for the tabletop scenario, as the problem is usually over-determined by the constraints. One can say that the Cartesian product grid is a compromise for QKD to ensure independence between users, but for a tabletop scenario where light sources are all in the same room, we observe that we can choose the decoy settings more arbitrarily (settings from different ports are allowed be correlated). For instance, we demonstrate that Latin hypercube sampling \cite{LHS} (LHS) method can be used to generate structured random decoy setting samples and drastically reduce the total number of settings needed (for instance from a minimum of 4096 for the full grid to as few as 256).

	\end{enumerate}

	In this paper, we will describe the Methods in Sec. II, including the setup for the sources, detectors, and circuit/channel, the linear regression training method, and the improved sampling strategy. In Sec. III we will show the numerical results of applying the trained model to various optical circuits/channels. We wrap up the paper in Sec. IV with some discussions and outlooks.

	\section{Methods}
	
	\subsection{Physical Setup and the Decoy-State Problem}
	
	Here we describe a general setup model for using coherent light to perform quantum information processing tasks. Let us consider some quantum circuits (or channels) that take in N inputs and N outputs, which we denote as $N \times N$ circuits/channels from here on. Ideally, we would like to study a circuit's performance if Fock states are fed in and photon-number-resolving (PNR) detectors are used to count the photons at each output port. Physically, though, we would like to use off-the-shelf phase-randomized coherent states at the input and threshold detectors at the output to simulate the effect of Fock states and photon counting detectors. We will consider three cases depending on our interest in the Fock-state photon statistics: full input/output statistics, input-only statistics, and hybrid statistics on selected ports.
	
	\begin{figure}[t]
		\includegraphics[scale=0.31]{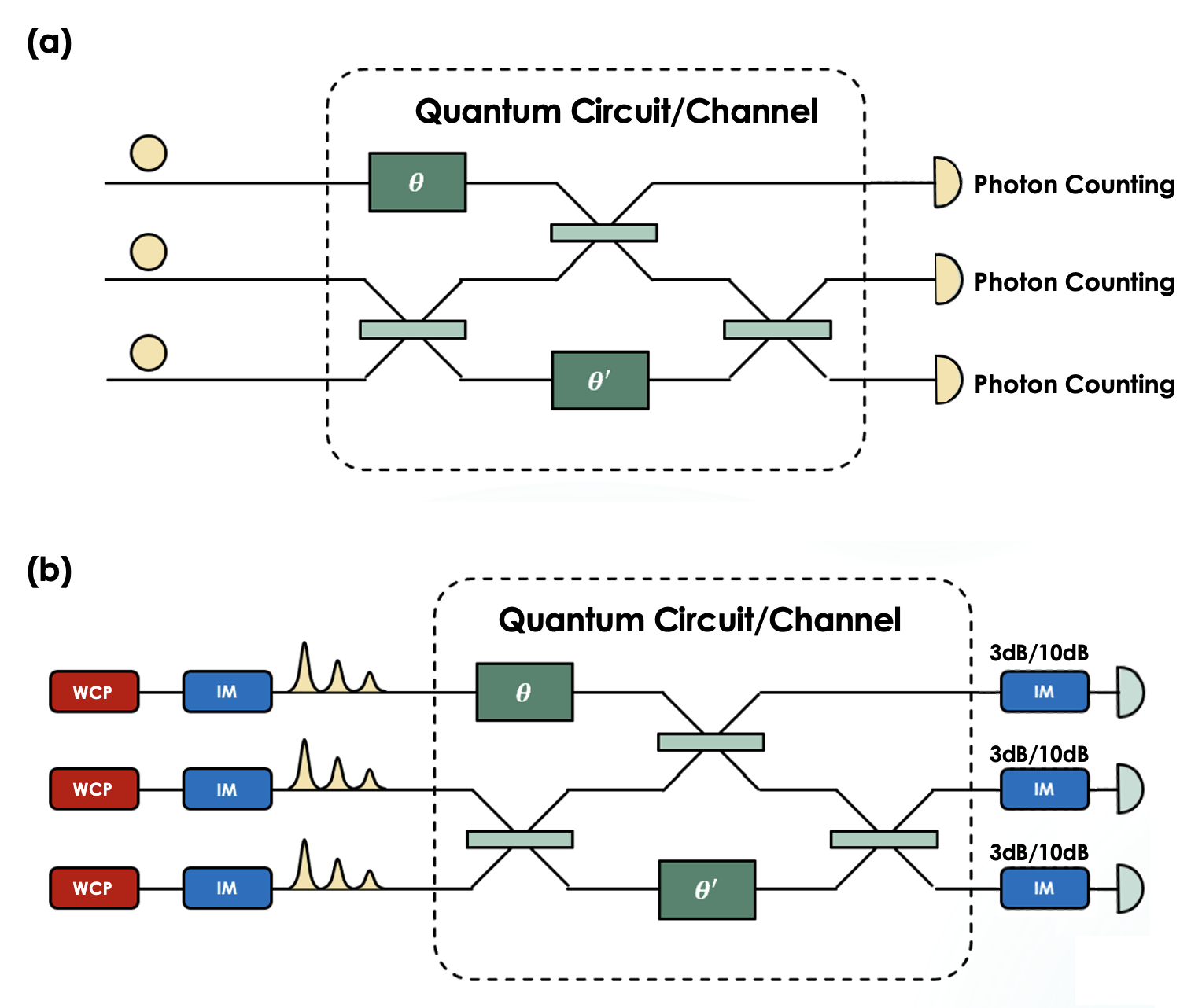}
		\caption{(a) A photonic information processing setup. The optical circuit usually needs Fock states as input and photon-counting detectors on the output to achieve desired function, such as optical quantum computing or quantum metrology. (b) A full input/output decoy-state setup where input Fock states are replaced with weak coherent pulses (WCPs) with varying intensities and the output detectors are replaced with threshold detectors with varying attenuations before them. Combined with post-processing (decoy state analysis), we can estimate the statistics coming from the events where desired photon numbers enter and leave the circuit in question.}
		\label{fig:1}
	\end{figure}
	
	1. We start with a full input/output decoy setup, where we are interested in the Fock state statistics across all input and output ports. An illustration can be found in Fig. \ref{fig:1}. This scenario is in fact the same as that of Ref. \cite{Marcos_interference}. For a general $N\times N$ circuit, suppose the input coherent states have intensities $\vec{\mu}=\{\mu_1,\mu_2,...\}$ where subscripts represent port index, the output detectors have attenuators $\vec{\eta}=\{\eta_1,\eta_2,...\}$ before them (detector efficiencies are incorporated into the attenuators too), and the target detection pattern is $\vec{p}=\{p_1,p_2,...\}$ (which is a length N array storing binary bits 1/0 signaling click/no-click). We denote the gain $Q_{\vec{\mu},\vec{\eta},\vec{p}}$ as the probability of detecting a pattern $\vec{p}$ given input/output settings of $\vec{\mu},\vec{\eta}$:
	
	\begin{equation}\label{eq:decoy_full}
		Q_{\vec{\mu},\vec{\eta},\vec{p}} = \sum_{\substack{\vec{n},\vec{m}}} P_{\vec{\mu}}(\vec{n}) D_{\vec{\eta},\vec{p}}(\vec{m}) Y(\vec{m}|\vec{n}),
	\end{equation}
	
	\noindent where the Poissonian coefficients are
	
	\begin{equation}
		P_{\vec{\mu}}(\vec{n}) = \prod_{i=1}^N e^{-\mu_i}{\mu_i^{n_i}\over {{n_i}!}},
	\end{equation}
	
	\noindent and the detector click coefficients are
	
	\begin{equation}
		D(\vec{\eta},\vec{p},\vec{m}) = \prod_{j=1}^N [p_j (1-(1-\eta_j)^{m_j}) + \overline{p}_j (1-\eta_j)^{m_j}],
	\end{equation}

	\noindent where $\overline{p}_j$ is defined as the logical negation of $p_j$, namely, $1-p_j \bmod 2$.

	To limit the dimensionality of the problem to a finite number, one can set a cutoff $S_n,S_m$ for the photon number in Eq. \eqref{eq:decoy_full} (using the fact that all $P(\vec{\mu},\vec{n})$ add up to 1):
	
	\begin{equation}\label{eq:decoy_full_cutoff}
		\begin{aligned}
			Q_{\vec{\mu},\vec{\eta},\vec{p}} &= \sum_{\substack{\vec{n} \in S_n \\ \vec{m} \in S_m}} P_{\vec{\mu}}(\vec{n}) D_{\vec{\eta},\vec{p}}(\vec{m}) Y(\vec{m}|\vec{n}) + \Delta_{\vec{\mu}},\\
		\end{aligned}
	\end{equation}

	\begin{equation}\label{eq:decoy_full_cutoff_2}
		\begin{aligned}
			0 &\leq \Delta_{\vec{\mu}} \leq \left(1 - \sum_{\vec{n} \in S_n} P_{\vec{\mu}}(\vec{n}) \right),\\
		\end{aligned}
	\end{equation}

	\noindent where we define the cut-off ranges $S_n,S_m$ for $\vec{n},\vec{m}$ in the sum. For instance, we can define rectangular bounding boxes of $n_i,m_j \leq s_{cut}$ or triangular bounding boxes of $\sum{n_i}, \sum{m_j}\leq s_{cut}$. Additionally, depending on the channel assumptions, we can add constraints, such as $\sum m_j = \sum n_i$ for an ideal lossless unitary circuit, or $\sum m_j \leq \sum n_i$ for a lossy unitary circuit. The condition also can be used for any general passive channel, which does not have to be unitary. The only condition is that there is no adversary injecting light into the detectors, which is reasonable for a tabletop setup in the lab – this is also the assumption in Ref. \cite{Marcos_interference}.\footnote{In the most general case, output photons only satisfy a global energy cap $\sum m_j \leq s_{cut}$ (i.e. allowing injected/stray photons at the detector). This generally makes the model very unstable due to including lots of high-order terms in the form of, for instance, $Y(5,5|1,1)$. Fortunately for us, in most of the scenarios we don't need photon counting on the receiver end of an open channel.}

	Given multiple settings of $\vec{\mu}$ and $\vec{\eta}$, commonly known as (input) decoy states and detector decoys, respectively, Eq. \eqref{eq:decoy_full_cutoff} can be formulated as a set of linear equations where $Q_{\vec{\mu},\vec{\eta},\vec{p}}$ are known observables, $P_{\vec{\mu}}(\vec{n}) D_{\vec{\eta},\vec{p}}(\vec{m})$ are known coefficients, and $Y(\vec{m}|\vec{n})$ (which remain unchanged across the constraints regardless of the $\vec{\mu}, \vec{\eta}$ settings) are the variables one would like to solve for. In simplified notations, we can reformulate the linear equations as

	\begin{equation}\label{eq:linear}
		\vec{Q} - \vec{\Delta}= A \vec{Y},
	\end{equation}
	
	\noindent where $\vec{Q}$ is the set of all observables iterating over all decoy settings of $\{\vec{\mu},\vec{\eta},\vec{p}\}$,  $\vec{\Delta}$ are the residue terms coming from Poissonian tails $\Delta_{\vec{\mu}}$, $A$ is the matrix containing all the coefficients $P_{\vec{\mu}}(\vec{n}) D_{\vec{\eta},\vec{p}}(\vec{m})$ (rows corresponding to decoy settings and columns corresponding to photon numbers), and $\vec{Y}$ is the flattened 1-D array of all variables/yields. We can see that this is a standard \textit{linear program}.

\begin{figure}[t]
		\includegraphics[scale=0.37]{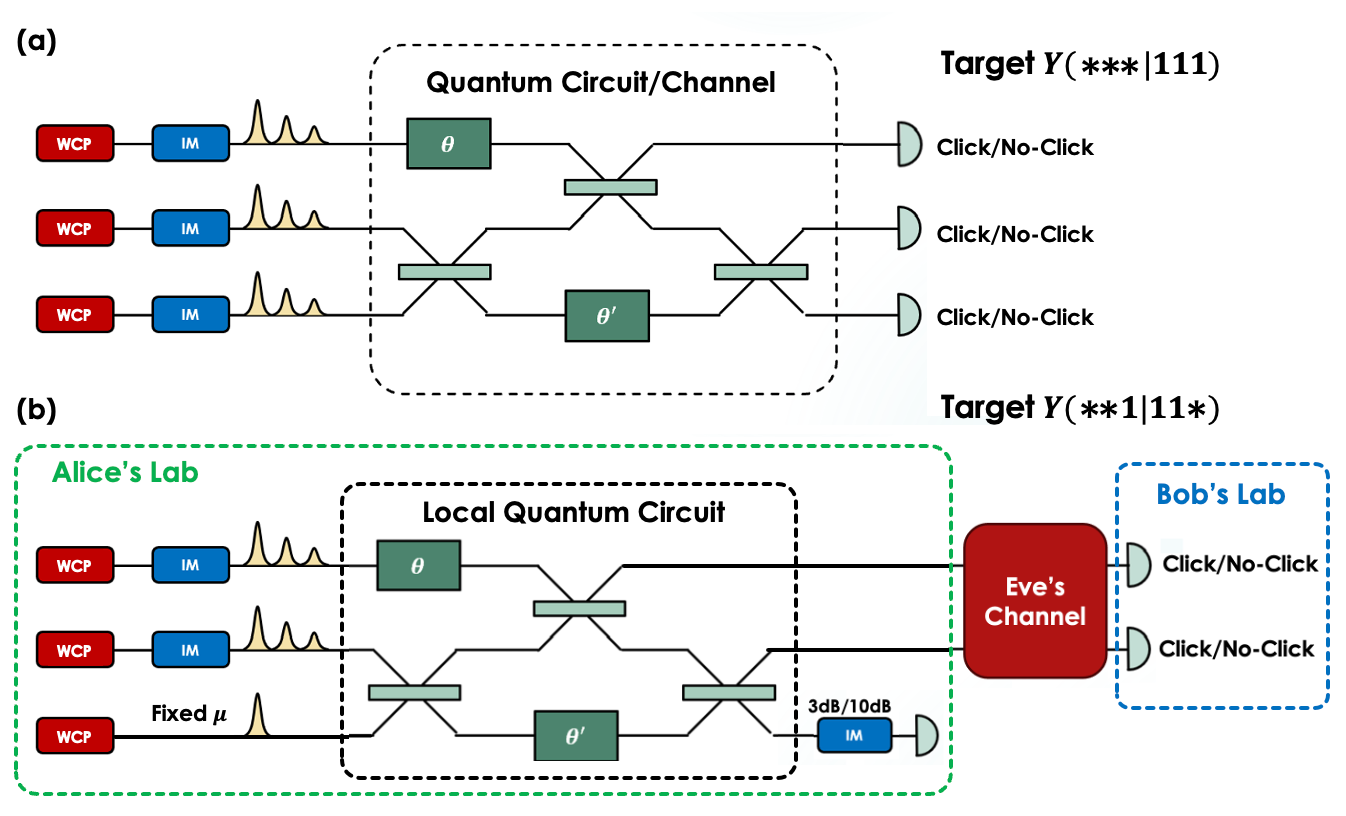}
		\caption{(a) A setup where only input ports carry decoy states while output ports have fixed detector efficiencies and only report click/no-click. In this case we are estimating the Fock state statistics for input ports only, $Y_{\eta_1,\eta_2,\eta_3,p_1,p_2,p_3}(\ast,\ast,\ast|n_1,n_2,n_3)$. (b) A setup with mixtures of ports with decoy states and fixed ports. For instance, here Alice can iterate over intensity settings for two ports and keep one port at constant (or vacuum) intensity. She can also choose to implement detector decoy on some output ports, such as for her local detectors for heralding, while keeping the other ports free-running, such as for ports going through an open quantum channel into Bob's detectors. In this case, we can estimate yields in the form of $Y_{\mu_3,\eta_1,\eta_2,p_1,p_2}(\ast,\ast,m_3|n_1,n_2,\ast)$}
		\label{fig:altsetup}
	\end{figure}

	2. In addition to the above full input/output decoy setup, it is also common (for instance in QKD, or in boson sampling with threshold detectors) to ignore output photon statistics and only record click/no-click on the threshold detectors. In this case, we can \textit{freeze} the output ports and do decoy-state analysis on only input ports:

	\begin{equation}\label{eq:decoy_input}
		Q_{\vec{\mu},\vec{\eta},\vec{p}} = \sum_{\substack{\vec{n}}} P_{\vec{\mu}}(\vec{n}) Y_{\vec{\eta},\vec{p}}(\vec{n}),
	\end{equation}

	\noindent where the observables are still the same set of $\{Q_{\vec{\mu},\vec{\eta},\vec{p}}\}$, but we group them in sets with the same $\vec{\eta},\vec{p}$ (or, more often, with only one set of fixed $\vec{\eta}$, since if the experiment does not involve detector-decoy, we do not need to vary output attenuations at all, in which case the observables are only grouped by detection pattern $\vec{p}$). We are therefore building linear equations across different sets of input intensities $\vec{\mu}$ to estimate input Fock statistics $Y_{\vec{\eta},\vec{p}}(\vec{n})$, the probability of observing $\vec{p}$ click pattern, given input Fock states $\vec{n}$. We can also denote the yield as $Y_{\eta_1,\eta_2,...,p_1,p_2,...}(\ast,\ast,...|n_1,n_2,...)$, where $\ast$ denote that we do not track the photon number in said port, i.e. decoy state method skips the ports denoted with $\ast$, and the obtained yield depends on a single specific set of output attenuator (detector efficiency) and pattern setting. See Fig. \ref{fig:altsetup} (a). This is still a linear program. Note that, to actually solve it, we would still need to define the cut-off ranges $S_n$ for $\vec{n}$ in the sum and bound the tail terms between 0 and 1, following a similar procedure as Eq. \eqref{eq:decoy_full_cutoff} to limit the number of variables.

	3. Moreover, we make the new observation that we do not even have to limit the decoy states to either full input/output or input/output-only. Rather, decoy-state analysis can be performed on arbitrary input ports and arbitrary output ports. The constraints can be written as:

	\begin{equation}\label{eq:decoy_arbitrary}
		Q_{\vec{\mu},\vec{\eta},\vec{p}} = \sum_{\substack{\vec{n_I},\vec{m_J}}} P_{\vec{\mu}}(\vec{n_I}) D_{\vec{\eta},\vec{p}}(\vec{m_J}) Y_{\vec{\mu}_{\overline{I}},\vec{\eta}_{\overline{J}},\vec{p}_{\overline{J}}}(\vec{m_J}|\vec{n_I}),
	\end{equation}

	\noindent where we define the tracked partial index sets $I,J \subseteq [N]$ (here $[N] = \{1,2,3,...,N\}$ is the full index set) for input and output ports, and $\overline{I}=[N]\backslash I,\overline{J}=[N]\backslash J$ are the complement sets (i.e. excluding $I,J$ indices). For example, if we track input ports $1,2$ and output port $3$ (as shown in Fig. \ref{fig:altsetup} (b)) in a $3\times3$ circuit, the sets would be $I=\{1,2\},J=\{3\}$ and the target variables $Y_{\vec{\mu}_{\overline{I}},\vec{\eta}_{\overline{J}},\vec{p}_{\overline{J}}}(\vec{m_J}|\vec{n_I})$ can be expanded as $Y_{\mu_3,\eta_1,\eta_2,p_1,p_2}(\ast,\ast,m_3|n_1,n_2,\ast)$, i.e. we combined the data from various $\mu_1,\mu_2,\eta_3,p_3$ to build a linear program, and we solve for the target yield that depend on a specific set of $\mu_3,\eta_1,\eta_2,p_1,p_2$. Again, to actually build the linear program, we need to set cutoffs $\vec{n_I}\in S_n,\vec{m_J} \in S_m$. We can also add the constraint $\sum m_j \leq \sum n_i$ if the circuit is passive and untracked inputs are vacuum states. This is a reasonable assumption if we use the setup like that of Fig. \ref{fig:altsetup} (b) for scenarios like heralded state preparation or even QKD, where untracked input ports are empty, the only photon counting happens in Alice's local heralding detectors, and the receiver end does not require photon counting. In this case, the passive circuit assumption is valid as long as the channel does not inject photons back into the source circuit.\\

	We would like to remark that all three above scenarios eventually all come down to the same-form linear program $\vec{Q} - \vec{\Delta}= A \vec{Y}$. The following techniques in our paper can be applied to all three scenarios (i.e. independent of the actual choice of input/output combinations of ports). As an example, in the following text we will mainly focus on scenario 1 (full input/output), where we solve for $Y(\vec{m}|\vec{n})$.
	
	Additionally, we note that scenario 3 is ideal in a quantum communication setup (such as QKD), but such an application requires provable upper/lower bounds on the Fock statistics, not just predictions, which is outside the scope of this paper that focuses on establishing the framework and performing accurate predictions. We will show such provable bounds in the companion paper, \cite{this_companion}, where scenario 3 leads to interesting use cases that combine local state preparation and remote quantum communication.

	For the above linear program, common LP solvers, such as Gurobi or Matlab linprog, can be used to help upper/lower bound each of the variables. In principle, it seems that the above Eq. \eqref{eq:decoy_full_cutoff} can theoretically bound $Y(\vec{m}|\vec{n})$ accurately up to any arbitrary photon number, so long as we have a sufficient number of signals for a large number of decoy settings. However, it turns out that this linear model is highly numerically unstable, due to two main reasons:

	\begin{itemize}
	
	\item The very large number of terms. For instance, a $3\times 3$ circuit with only 2 decoy states and 2 detector decoy settings per port will have a total of $2^6$ settings and $2^3$ detector patterns, summing up to as many as $512$ constraints, and photon cutoff of merely $S=5$ will also result in about $10000$ variables, resulting in a large coefficient matrix size of $512 \times 10000$;\\
	
	\item The discrepancies in orders of magnitude of coefficients. For instance, $Y(1,1,1|1,1,1)$ has a coefficient on the order of $10^{-3}$, compared to say $Y(5,5,5|5,5,5)$ that is present in the same constraints but having a coefficient on the order of $10^{-11}$. This makes the matrix $A$ fundamentally ill-conditioned.\\
	
	\end{itemize}

	Because of the above numerical instability, directly solving Eq. \eqref{eq:decoy_full_cutoff} will result in very limited maximum achievable photon numbers and circuit dimensions. For instance, in Ref. \cite{Marcos_interference}, it was only possible to do a (noisy) estimation up to $Y(3,3|3,3)$ for a $2\times 2$ beamsplitter and $Y(1,1,1|1,1,1)$ for a $3\times 3$ tritter. In the next section, we will instead discuss a linear regression approach to enable higher accuracy and larger dimension.

	\subsection{Training a Linear Regression Model}

	In this section, we propose to use a linear regression method to learn the mapping from coherent state observables to the Fock state statistics we want. 
	
	Here we consider the problem: Is it possible to obtain the target $Y(\vec{m}|\vec{n})$ (i.e. an element of $\vec{Y}$) from the set of observables $\vec{Q}=\{Q_{\vec{\mu},\vec{\eta},\vec{p}} \}$, without explicitly solving the decoy state model $\vec{Q} - \vec{\Delta}= A \vec{Y}$ as shown in Sec. II.A?
	
	Intuitively, we observe that, if the photon number cutoff is large enough, 
	
	\begin{equation}\label{eq:linear_approx}
		\vec{Q} \approx A \vec{Y}.
	\end{equation}
	
	Therefore, the process of either solving the linear program or finding a mapping by machine learning is fundamentally solving the matrix inversion\footnote{A is a square matrix only if the number of constraints equals variables. We can always choose the cutoff boundary or choose sets of decoy settings to achieve this. If A is for instance over-conditioned, though, the inversion becomes a pseudo-inversion, say ``pinv" in Matlab, which is the Moore-Penrose pseudo inversion.}
	
	\begin{equation}\label{eq:inverse}
		\vec{Y} \approx A^{-1} \vec{Q}.
	\end{equation}

	\noindent It is just that, since A is an extremely ill-conditioned matrix (due to the input Poissonian structure combined with output photon statistics), its inverse is very difficult to find. Nonetheless, this actually hints that each row has a linear structure $Y(\vec{m}|\vec{n}) = \sum_i c_i Q_i$.\footnote{We can train the linear regression model in the standard form $Y(\vec{m}|\vec{n}) = \sum_i c_i Q_i + b$, but for each particular $\vec{\mu},\vec{\eta}$, we always have a set of $\sum_{\vec{p}} Q_{\vec{\mu},\vec{\eta},\vec{p}} = 1$. This means that any constant intercept $b$ can be rewritten as $b \times \sum_{\vec{p}} Q_{\vec{\mu},\vec{\eta},\vec{p}}$ and assimilated into the coefficients of one (or more) set of observables, as long as there is a full combination of patterns.} Also, note that the coefficients $\{c_i\}$ should be independent of the channel itself and are only related to the decoy state settings, i.e. the elements in A itself. This means that there should be a linear mapping $\{Q_{\vec{\mu},\vec{\eta},\vec{p}} \} \rightarrow Y(\vec{m}|\vec{n})$ that points from the coherent-state observables to the Fock state statistics that we want, and this mapping is only determined by the set of $\{\vec{\mu},\vec{\eta},\vec{p}\}$ we choose before an experiment, \textit{independent} of the actual channel itself.
	
	We propose to apply a linear regression model to \textit{learn} the mapping $\{Q_{\vec{\mu},\vec{\eta},\vec{p}} \} \rightarrow Y(\vec{m}|\vec{n})$. Such an idea of using supervised learning to learn a fixed but difficult-to-compute (closed form unknown) function/mapping was proposed to predict optimal QKD parameters in our earlier paper \cite{ML1} in 2019 (along with a concurrent work Ref. \cite{ML2} by Lu et al., as well as many follow-up works in the community, such as Refs. \cite{ML3,ML4,ML5,ML6,ML7}). Here we use a similar train of thought to help us predict the Fock state statistics in decoy state analysis.
	
\begin{figure}[t]
		\includegraphics[scale=0.55]{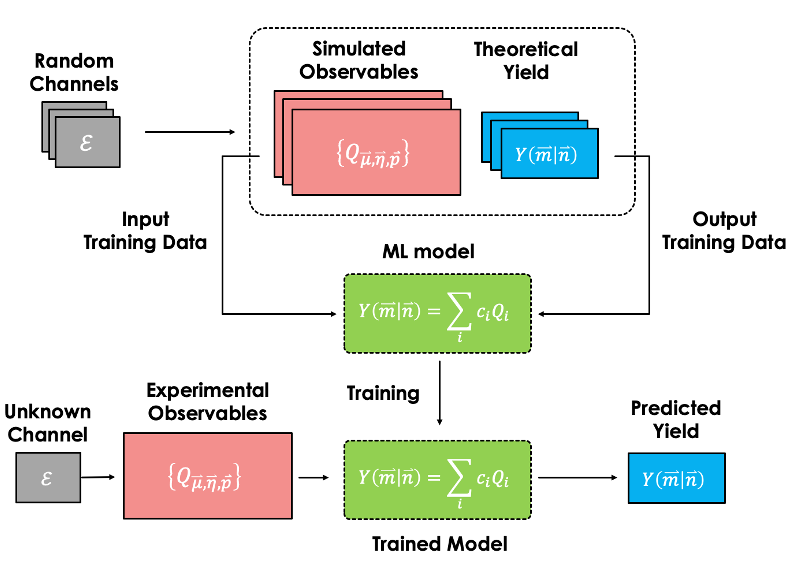}
		\caption{The data flow for the training. We can generate random channels (arbitrary or unitary, depending on target user scenario) and calculate the simulated observables and theoretical yield for each one, which can then be used as labeled data to train a machine learning model (which can be as simple as a linear regression model) to learn the mapping. Importantly, this mapping is channel-independent and is determined only by the decoy settings chosen prior to an experiment. Once the model is trained, it can be used to read in new observables from coherent light sources obtained from an unknown channel and predict the corresponding Fock state yield with high accuracy.}
		\label{fig:fig_ML}
	\end{figure}

	As shown in Fig. \ref{fig:fig_ML}, we train it with pairs of labelled data $[\{Q_{\vec{\mu},\vec{\eta},\vec{p}} \}, Y(\vec{m}|\vec{n})]$ generated from random channels, and once the linear mapping is learned, it can be used to read in observables $\vec{Q}$ obtained from a new unknown channel and predict its correct Fock state statistics. 

	To implement the training, we need to select a particular type of channel model. The learnt mapping will only work on the same type of channel used for training. Here we consider two scenarios:

	\begin{enumerate}
		\item For the most general case of a passive quantum channel diagonalized in Fock space (i.e. with phase randomization), we simply allow each particular $Y(\vec{m}|\vec{n})$ to be a random number within $[0,1]$. The observables would simply be calculated from Eq. \eqref{eq:decoy_full_cutoff} (up to a cutoff photon number and omitting the negligible Poissonian tail). The randomly generated set of $Y(\vec{m}|\vec{n})$ are normalized to ensure $\sum_{\vec{m}} Y(\vec{m}|\vec{n}) = 1$.
		\item  For many user scenarios, as mentioned in Sec. I, the circuits are actually unitary with known Hilbert space dimensions (despite that we may not know the exact circuit to be characterized). In this case, we simply draw $U$ from Haar-random unitaries and use it for training.
		
	\end{enumerate}

	For the unitary case, given input coherent states of intensities $\vec{\mu}$, we can equivalently describe the input states by their complex amplitudes $\vec{\alpha}$ where $\alpha_i = e^{i\phi_i} \sqrt{\mu_i}$. The output coherent states after passing through the circuit become
	
		\begin{equation}\label{eq:coherent_unitary}
			\vec{\beta} = U \vec{\alpha},
		\end{equation}
		
		\noindent which further triggers each detector at output port $j$ to fall into click pattern $p_j$ with probability
		
		\begin{equation}
			P_j = p_j[1 - e^{-\eta_j|\beta_j|^2}] + \overline{p}_j e^{-\eta_j|\beta_j|^2},
		\end{equation}
		
		\noindent so we can easily calculate the gain (given fixed input phases) based on
		
		\begin{equation}
			Q_{\vec{\mu},\vec{\eta},\vec{p},\vec{\phi}} = \prod_j P_j,
		\end{equation}

		The final gain is a phase-randomized integration over all input phases in $[0,2\pi)$:

		\begin{equation}
			Q_{\vec{\mu},\vec{\eta},\vec{p}} = {1\over{(2\pi)^N}}\int_{\phi_1=0}^{2\pi} ...  \int_{\phi_N=0}^{2\pi} Q_{\vec{\mu},\vec{\eta},\vec{p},\vec{\phi}} d\phi_1 ... d\phi_N
		\end{equation}

		\noindent which, in simulation, can be calculated with discrete and uniform phase sample points for this multivariate integration in all $\phi_i$ dimensions. Heuristically, we use small values (such as just 5 points) on the circumference, which enables the efficient generation of large batches of data while already sufficing in calculating $Q_{\vec{\mu},\vec{\eta},\vec{p}}$ with little error. Equivalently, this also means that if we are to probe an unknown circuit, experimentally we can choose to prepare the pulses with discrete phase randomization (such as commonly done for QKD \cite{discrete_phase_randomization}), which will result in almost-identical gains as the continuously phase-randomized ones and hence obtain almost-identical predictions of yields.
		
		Meanwhile, it is well-known that the photon yields, $Y(\vec{m}|\vec{n})$, are proportional to the squared permanent of the unitary matrix $U$ (more precisely, its related matrix $U'$) \cite{permanent1,permanent2}:
		
		\begin{equation}\label{eq:permanent}
			Y(\vec{m}|\vec{n}) = \left| {{\Perm(U')}\over{\sqrt{(\prod m_j!)(\prod n_i!)}}} \right|^2,
		\end{equation}
		
		\noindent where $\Perm$ is the matrix permanent, and $U'$ is a matrix derived from $U$ by repeating the $i$-th column for $n_i$ times and $j$-th row by $m_j$ times. For instance, in the case where all elements of $\vec{n}$ and $\vec{m}$ are 1, the yield $Y(1,1,..,1|1,1,..,1)$ is proportional to the square of the permanent of $U$ itself.
		
		We make a key observation: \textit{Each yield $Y(\vec{m}|\vec{n})$ is not an independent variable, but is determined by the elements of matrix $U$ itself.} This means that there are much fewer unknown degrees-of-freedom in the problem - in fact only $N\times N$ complex numbers, with the additional unitary constraint. Thus, directly optimizing the linear program in Eq. \eqref{eq:decoy_full_cutoff} is a highly \textit{pessimistic} approach since it assumes all variables are independent. In a way, this is like a unitary learning problem, but our goal is even simpler than obtaining elements of $U$, since we are only interested in one single attribute of it, which is $Perm(U')$. 
	
	For the original LP problem, the matrix $A$ defined in Eq. \ref{eq:linear} is very ill-conditioned because $\det A$ approaches $0$ exponentially fast as the dimension of $A$ increases, which depends on the decoy setting number and photon number cutoff, making the LP model very difficult to solve. On the other hand, if we consider the inverse matrix $A^{-1}$, its rows and columns are in fact made up of alternating positive and negative elements, some with very large amplitudes. In other words, we need to add/subtract elements in $\vec{Q}$ with alternating positive and negative coefficients that largely cancel out with each other, while the output depends on the differences between large numbers. Therefore, evaluating Eq. \eqref{eq:inverse} according to the book can suffer serious rounding error\footnote{In fact, the task itself of finding a strategy to minimize the rounding error when summing large positive/negative numbers is an NP-hard problem \cite{rounding_error}.}. In addition to rounding error, statistical fluctuations and experimental measurement inaccuracies will also result in disturbances to the observables $\vec{Q}$, which makes the calculation of $A^{-1} \vec{Q}$ further unstable, in particular for higher order terms.

	As an alternative intuitive explanation, for an arbitrary non-unitary scenario, we note that $\left|{\partial Q_i}\over{\partial Y(\vec{m}|\vec{n})}\right|$ can be extremely small. For instance, the yield $Y(5,5,5|5,5,5)$ will have negligible impact on $Q_i$ due to the exponentially small Poissonian coefficients. In other words, $\left|{\partial Y(\vec{m}|\vec{n})}\over{\partial Q_i}\right|$ is very large for high-photon-number terms, so the estimated yield will be extremely sensitive to small disturbances in observed gains. This makes the estimation of higher order terms very unstable in the presence of statistical fluctuations, experimental measurement inaccuracies, and limited machine precision (rounding error). Such an effect fundamentally limits the maximum accuracy and circuit size in the scenario where no assumption is made on the channel, regardless of whether linear programming or the new linear regression method is used.
	
	For the unitary scenario, on the other hand, in fact the yields not only satisfy the expression in Eq. \eqref{eq:decoy_full_cutoff}, but also are subject to Eqs. \eqref{eq:coherent_unitary} - \eqref{eq:permanent}, so each yield term is not independent. It does not suffer from the issue of $\left|{\partial Q_i}\over{\partial Y(\vec{m}|\vec{n})}\right|$ being small, since when any specific $Y(\vec{m}|\vec{n})$ changes, every other yield also must change simultaneously (since it means the unitary $U$ is changing). This makes it possible to train a linear regression model for the unitary scenario efficiently. See Sec. III for numerical results and applications.

	By the way, we note that Refs. \cite{Xiongfeng_interference,Chau_Vandermonde,Hayashi_Vandermonde} showed that one can analytically compute the inverse for a Vandermonde matrix, which is closely related to the coefficient matrix $A$ in the decoy state method. Ref. \cite{Xiongfeng_interference} also discussed this in the context of multiport circuits, although its approach only worked for input decoys. In Appendix B, we demonstrate that we can build on this approach and extend it to a hybrid analytical-numerical method (analytical inverse for input decoys, numerical optimization for output decoys) that works for more general cases of mixed input and output decoys. However, such a method also makes no assumption about the channel and suffers from the same instability for higher order term recovery. Compared with the linear regression method, we found that it always requires more decoy settings and yields less accurate results (despite having the advantage of requiring no pre-training). Therefore, in this paper we adopt the linear regression method as our approach and present it as the main result.

	\subsection{Sampling Strategies}
	
	\begin{figure}[h]
		\includegraphics[scale=0.35]{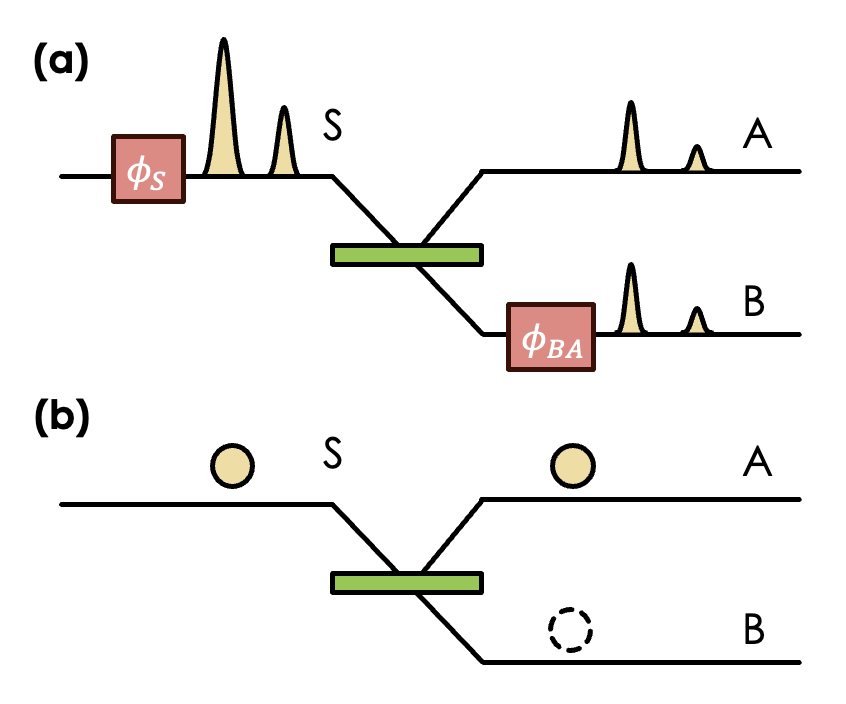}
		\caption{(a) Synchronized intensity choices at ports A and B can be viewed equivalently as intensity choices at a common virtual source S (phase-randomized by a global phase $\phi_S$) and a randomized relative phase $\phi_{BA}$ between A and B, which does not affect photon numbers. Therefore, as shown in (b), performing decoy analysis with intensity settings of fixed ratios across ports can only tell the Fock state statistics of $n$ photons at port S, in other words a \textit{total} of $n_A+n_B=n$ photons in ports A and B, but not the specific Fock state statistics for $n_A,n_B$.}
		\label{fig:sync}
	\end{figure}

	\begin{figure}[h]
		\includegraphics[scale=0.35]{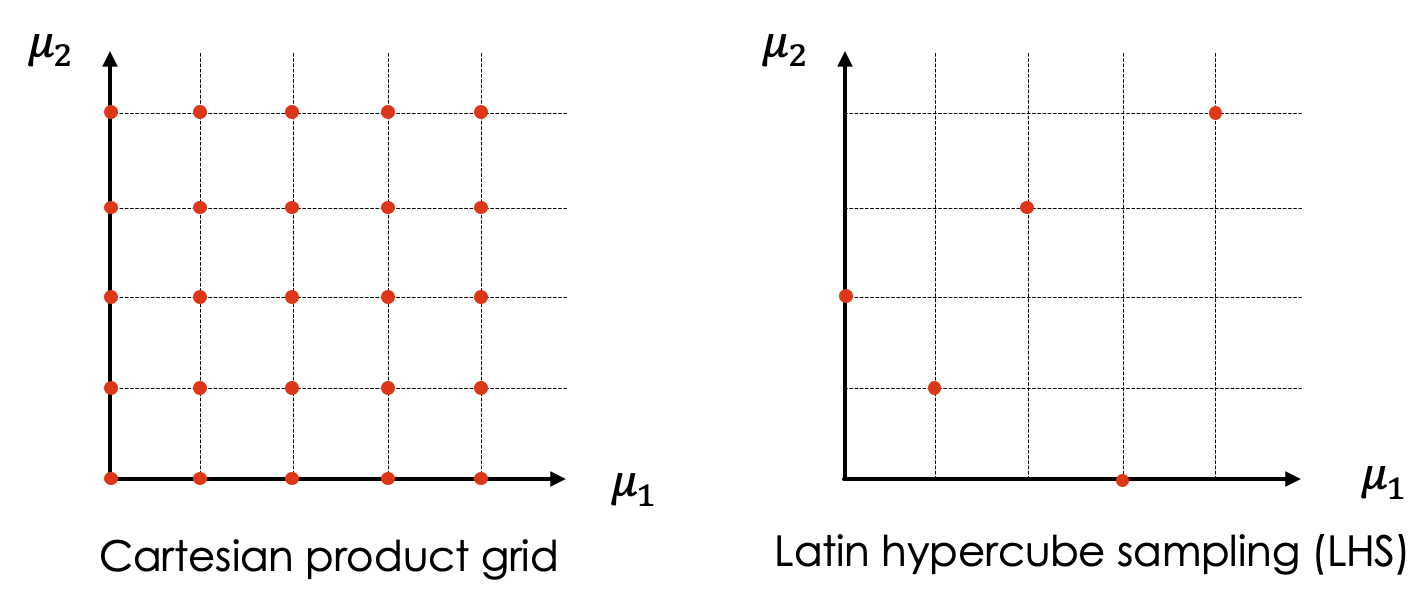}
		\caption{Full Cartesian product grid (left) versus Latin Hypercube Sampling (right). LHS is akin to the "8 rook problem" in chess (place every rook piece such that each row and each column contains exactly one rook), which creates a structured pseudo-random sampling across all dimensions with drastically smaller sample size.}
		\label{fig:LHS}
	\end{figure}

	In this section we define an additional important technique to reduce the data source needed and enable larger-scale circuits with the decoy state plus linear regression approach, namely by choosing a better sampling strategy for the decoy settings.

	In QKD protocols with multiple users (such as MDI-QKD or CKA), the users are required to each choose their decoy settings independently (or else the choice would cost them pre-shared random bits). Therefore, the sampling strategy is usually a full Cartesian product grid. For instance, for MDI-QKD, if the intensity set is $\{\mu,\nu,\omega\}$ for each of Alice and Bob, then the two users are expected to choose their setting $[\mu_1,\mu_2]$ pair from $\{[\mu,\mu],[\mu,\nu],[\mu,\omega],[\nu,\mu],[\nu,\nu],[\nu,\omega],[\omega,\mu],[\omega,\nu],[\omega,\omega]\}$. For a bigger $N\times N$ circuit, however, the number of required settings for such a full grid will soon blow up. For instance, for a $6\times 6$ circuit, even if we only use two intensity settings for each input port, two attenuator settings for each output port, and record all output patterns, we would have $2^6 \times 2^6 = 4096$ decoy settings and a total of $4096 \times 2^6 = 262,144$ observables, which is simply impractical as each observable would suffer from huge statistical fluctuations within any reasonable collection time.

	For our purpose (characterizing a circuit in the lab), we make the observation that the ports do not have to be independent, i.e. we can pick correlated sample points for the intensities, since a single user (Alice) determines the decoy settings across all ports, and there is no need at all to use a full Cartesian product.

	However, we cannot simply use locked intensity ratios (i.e. synchronized settings) across ports, such as using only the diagonal elements $\{[\mu,\mu],[\nu,\nu],[\omega,\omega]\}$. This is because, as shown in Fig. \ref{fig:sync}, decoy settings of locked intensity ratios will only tell us the total photon number across the synchronized ports, but not the Fock state statistics for each individual port.

	To alleviate this problem, we propose to use the Latin Hypercube Sampling (LHS) method \cite{LHS}, as shown in Fig. \ref{fig:LHS}, which is a pseudo-random (structured random) sampling method aimed at reducing the correlation between settings and allowing each dimension to be sampled across all discretized grid points exactly once. As we will later show in the numerical results, using LHS drastically reduces the number of samples needed. For instance, it can reduce 256 decoy settings to just 32 settings in a $4 \times 4$ circuit, or from 4096 decoy settings to just 256 settings for a $6 \times 6$ circuit.

	Note that, however, we are not claiming constant scaling for the resources. This is because Eq. \eqref{eq:coherent_unitary} requires only polynomial complexity to compute, but the calculation of permanent itself still has exponential complexity. This gap means that one must ``pay the price" in the sampling complexity (i.e. the overall scaling of the number of samples needed would likely still be exponential). However, at small to medium-scale circuit sizes, LHS offers us a much better constant factor despite the scaling, for instance making circuits of size $6\times 6$ still computable and feasible within reasonable experiment time.

	\section{Numerical Results and Applications}

			\begin{figure}[h]
		\includegraphics[scale=0.75]{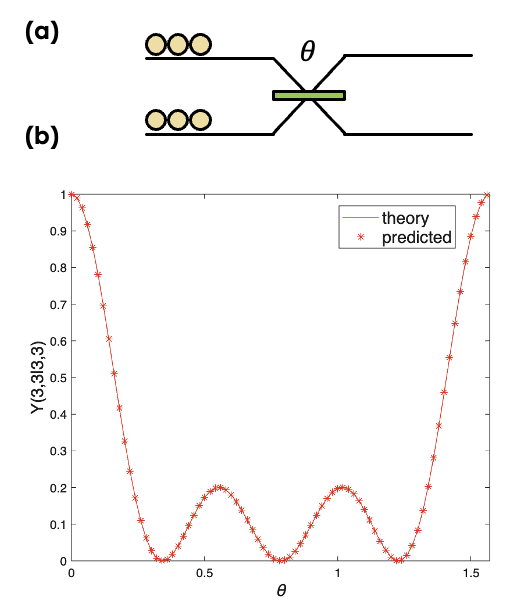}
		\caption{(a) A simple $2\times 2$ circuit with a beamsplitter of varying transmittance controlled by $\theta$. (b) Estimation of $Y(3,3|3,3)$. The linear regression model predicts the yield precisely with negligible noise, using only two intensity settings $\{0.5,0.1\}$ at each input port and two detector attenuation settings $\{0.1,0.5\}$ at each output port.}
		\label{fig:Y33}
	\end{figure}
	
	\begin{figure}[h]
		\includegraphics[scale=0.6]{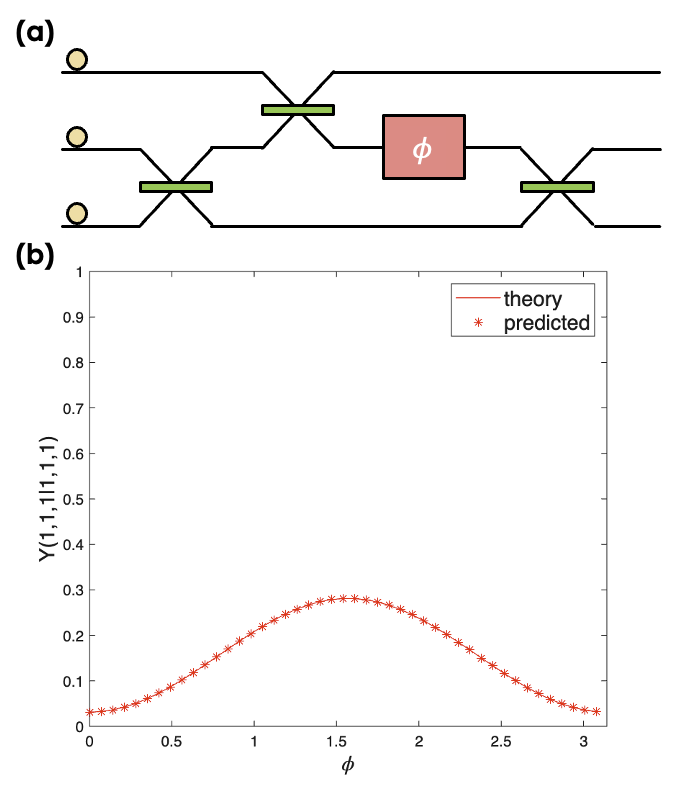}
		\caption{(a) A $3\times 3$ circuit with fixed 50-50 beamsplitters and a phase shifter $\phi$. (b) Estimation of $Y(1,1,1|1,1,1)$. The linear regression model, again, predicts the yield perfectly, using intensity settings $\{0.5,0.1\}$ and detector attenuation settings $\{0.5,0.1\}$ for each port.}
		\label{fig:Y111}
	\end{figure}

		\begin{figure}[t]
	\includegraphics[scale=0.45]{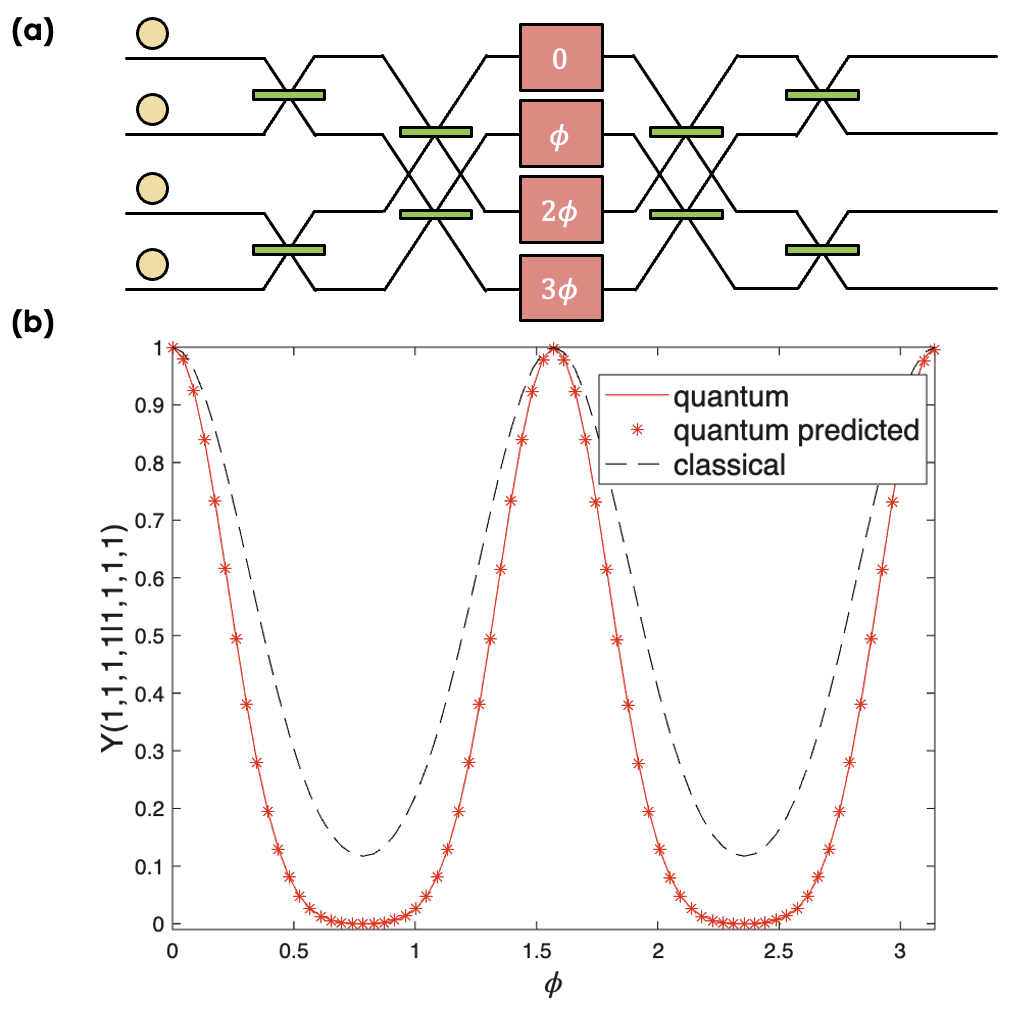}
	\caption{(a) The quantum metrology (phase detection) circuit in Ref. \cite{metrology1}. The unknown $\phi$ can be detected using four-photon inputs with higher sensitivity than classically achievable. (b) Simulated data for $Y(1,1,1,1|1,1,1,1)$ for ideal input photons, for distinguishable photons (i.e. classical case), and predicted with the linear regression model from WCP sources. The prediction is extremely accurate even with only 32 Latin Hypercube sampling points, and such prediction of Fock state statistics allows us to observe the higher quantum-only visibility. Note that for points with very small true value, such as at the bottom of the dips where $Y\approx 0$, the predicted values may occasionally be slightly smaller than zero, due to an absolute error noise level (on the order of $10^{-4}$) coming from using LHS, in which case we should truncate the occasional negative prediction results to zero. If one wishes to obtain predictions with no negative points, one can alternatively use the Cartesian product grid here, which contains just 256 points and can offer near machine precision with an absolute error on the order of $10^{-15}$ if observed values are noiseless.}
	\label{fig:metrology}
\end{figure}

\begin{figure}[h]
	\includegraphics[scale=0.45]{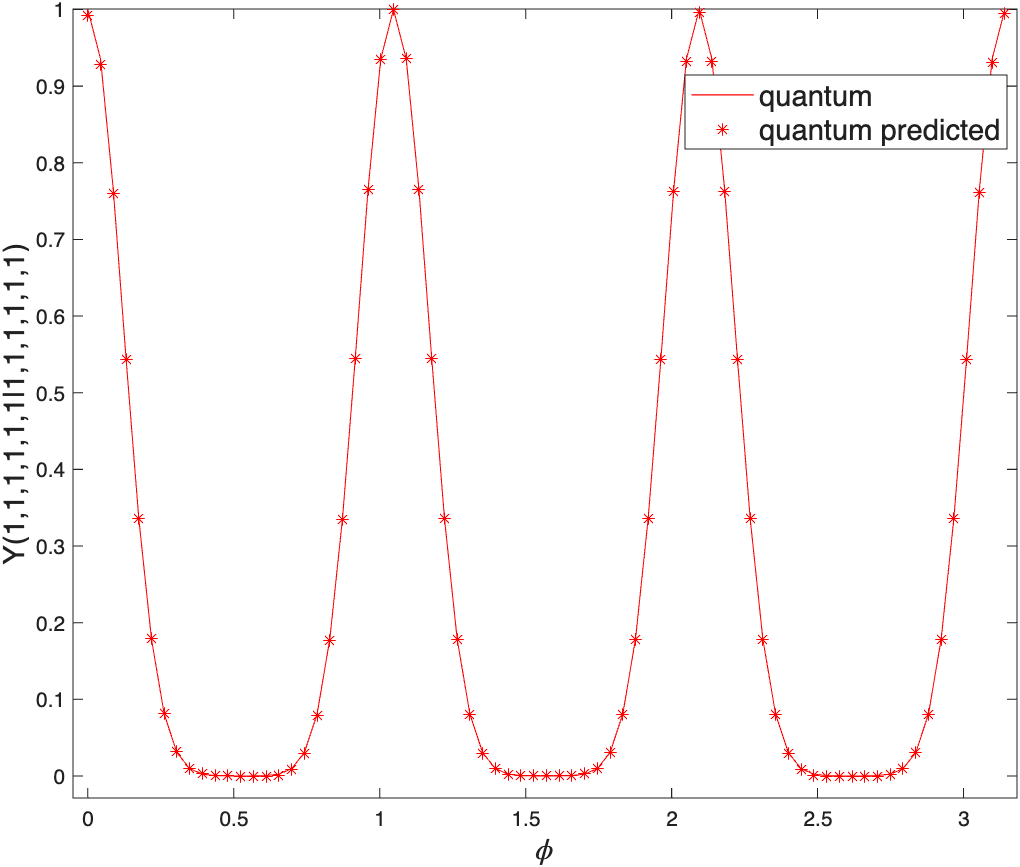}
	\caption{Simulation for a generalization of the phase detection circuit (phase values sandwiched between a quantum Fourier transform and its inverse) in Ref. \cite{metrology1} to $6\times 6$. We show that we can accurately predict $Y(1,1,1,1,1,1|1,1,1,1,1,1)$ with the linear regression model using 256 LHS points. Similarly, at the bottom of the dips where $Y\approx 0$, the predicted values may occasionally be slightly smaller than zero and need truncation due to the absolute error noise floor.}
	\label{fig:metrology2}
\end{figure}

Here we first begin by testing the above approaches with simple circuits. For instance, a $2\times 2$ unitary circuit and a $3\times 3$ unitary circuit. For small scale circuits like this, a full Cartesian product grid suffices. As shown in Figs. \ref{fig:Y33} and \ref{fig:Y111}, given the unitary assumption, the machine learning (ML) approach using a simple linear regression model almost always predicts yields perfectly.

We can then test with some more complicated circuits. For instance, we pick the four-photon interference setup in Ref. \cite{metrology1} as an example and try to reproduce the results with coherent light alone + decoy state method. Here we input phase randomized coherent lights either using a full Cartesian grid (with two possible decoy settings $\{0.5,0.1\}$ for $\mu_i$ at each port and set attenuators to also two possible values $\{0.5,0.1\}$ for $t_j$ at each port) or with Latin Hypercube sampling. As can be seen in Fig. \ref{fig:metrology}, the linear regression model can predict $Y(1,1,1,1|1,1,1,1)$ with pinpoint accuracy, which is much better than a linear programming model can achieve – in fact an LP solver can at most solve $Y(1,1,1|1,1,1)$ for a tritter at the expense of numerical noises and cannot solve $Y(1,1,1,1|1,1,1,1)$ at all. We even further extended the phase detection circuit (which is based on a Quantum Fourier Transform) in \cite{metrology1} to $6\times 6$ and tested $Y(1,1,1,1,1,1|1,1,1,1,1,1)$, and we can see in Fig. \ref{fig:metrology2} that the linear regression works in tracking as many as 6 photons.

Note that here we are achieving something remarkable - we are \textit{overcoming limitations on phase resolution for classical light using classical light} - just with the help of post-selection alone. In other words, we are post-selecting quantum events out of an apparently classical mixture. Moreover, in fact Ref. \cite{metrology1} only utilizes non-entangled single photons (whose indistinguishability results in the quantum advantage), but it is also possible to apply our framework to circuits that create, for instance, heralded EPR-pairs or even NOON-states for a bigger quantum advantage \cite{NOON}, but implementing them with simple WCP sources combined with post-selection.

Additionally, as mentioned in Sec. II.B, we can also switch the training data from random unitary circuits to non-unitary (say lossy) circuits or even arbitrary quantum channels with randomized yields. The only requirement for the channel is that it must be passive, i.e. no more outgoing photons than input photons. This would be suitable for a non-adversarial setup, such as a lossy or even non-linear local circuit, or an adversarial setup with only local ports photon-number-resolved, such as a heralded state preparation setup with an open quantum channel and threshold detection at the receiver.

For instance, we can test a 4-input, 4-output quantum channel and again train the linear regression model to find $Y(1,1,1,1|1,1,1,1)$. For unknown channels, since we no longer have the physical prior knowledge that locks higher-order terms to lower-order terms, we need more decoy intensity settings to obtain enough information to bound higher-order terms, for instance $\{0,0.05,0.3,0.7\}$ for each input port, while we keep output decoy settings as the same $\{0.5,0.1\}$ for each output port. In Table I, we can see that linear regression still can offer very high precision despite not having the unitary physical prior. Again, such a number of observables, $4^4 \times 2^4 \times 2^4$ for $\vec{u},\vec{t},\vec{p}$, significantly exceeds what linear program solvers can deal with.

Lastly, in Appendix A, we further discuss some practical imperfections, including circuit loss, statistical fluctuations, and intensity fluctuations. We show that our method (in particular the known-dimension circuit scenario with a physical prior) performs well even in the presence of imperfections, i.e. the approach is highly practical.

\begin{table}[htbp]
	\label{table:random_channel}
	\centering
	\caption{Examples of predicted data versus true yields for randomized passive 4-input-4-output channels. The full Cartesian product grid is used here. Data is presented with 5 decimal places and the absolute error is rounded to 2 significant digits. The average absolute error (over a testing set of 100 random channels) is $2.7\times 10^{-5}$.}
	\begin{tabular}{ccc}
		$Y_{\text{true}}$ & $Y_{\text{regression}}$ & Abs. Error\\
		\hline
		$0.01918$ & $0.01919$ & $5.2\times 10^{-6}$\\
		$0.00212$ & $0.00215$ & $3.0\times 10^{-5}$\\
		$0.00510$ & $0.00511$ & $1.7\times 10^{-6}$\\
		$0.02200$ & $0.02202$ & $2.9\times 10^{-5}$\\
		$0.02746$ & $0.02743$ & $3.2\times 10^{-5}$\\

	\end{tabular}
\end{table}

\section{Discussions}
	
	In this manuscript, we study the framework of using coherent light sources and threshold detectors, combined with decoy states and detector decoys, to effectively simulate Fock state sources and photon-counting detectors. We consider a general framework of decoy state method with arbitrary tracked input/output ports, and we propose a machine learning-based estimation tool (which works for arbitrary channel types, but works especially well for unknown unitary circuits) that can predict photon state statistics with much higher accuracy, acceptable photon number, and dimensions than prior art methods. We showed the effectiveness of our method for unitary circuits (such as those used in quantum metrology) and random channels, and we show that our method is robust in the presence of circuit loss, statistical fluctuation, and intensity fluctuation.
	
	Furthermore, if some output ports are local and some are open (released to an external channel), one can use our method to prepare heralded complex quantum states beyond single photons or just pairs of single photons. In our companion paper \cite{this_companion}, we will further show how to use this framework and obtain certified security bounds for the predicted values, enabling the use of our method for quantum communication/cryptographic applications.

	One additional note for the quantum metrology application discussed above: we are not claiming a fundamentally higher Fisher information \textit{per pulse} over classical resources, due to the heavy post-selection that decoy-state method entails. Using our method, the post-selected signals will recreate near-identical click statistics as a real quantum state (and all the high-visibility, high-sensitivity advantages), but the amount of total data sampled will dilute such advantage. This is perhaps a similar dilemma faced with probabilistic creation of certain quantum states (say a NOON state) with current technologies, unless heralding and active feed-forwarding is enabled to let only successful events through. Therefore, the main application of the decoy-state method, we think, is to be able to let user build a testing ground to, for instance, calibrate metrology detection setups or characterize optical circuits/gates \textit{offline}, as well as to allow them to create complex quantum states using post-selection, such as single photons, NOON states, EPR pairs, and GHZ states, etc., with easy-to-access components in any lab.\\
	
	\section{Acknowledgements}
	
	We thank Jinjie Li, Koji Azuma, William Munro, David Elkouss, Daniel Oblak, Ashutosh Singh, Barry Sanders, Stefan Scheel, Alexander Szameit, Johannes Bentzien, Urbasi Sinha, Norbert Lutkenhaus, Shlok Nahar, Lars Kamin, John Burniston, Masahiro Hotta, and Michiko Inoue for constructive discussions and helpful suggestions. W Wang acknowledges support from the JSPS Grant-in-Aid for Early-Career Scientists (KAKENHI Wakate) Grant, the University of Calgary and Quantum City start-up fund, and the NSERC Discovery Grant.

\appendix

\section{Robustness against Imperfections and Noises}

In this section, we discuss the practical imperfections and noises our setup may encounter and demonstrate its robustness in the presence of such factors.

\subsection{Lossy Circuits}

Firstly, for physical optical circuits, it is impossible to achieve perfect unitarity, since circuits/components themselves have internal losses, and connecting light sources and modulators to the circuits in question results in insertion losses. In the case of known losses, it is in principle possible to assimilate it into the source decoy intensity or detector efficiencies (i.e. calculate the yield of a ``virtual" lossless circuit), but there may be cases where we would like to calculate the actual input-output statistics of a given physical circuit and not its ideal lossless portion, or cases where there are internal losses or unknown imbalanced losses across the ports which cannot be easily assimilated into sources/detectors. Therefore, it is important for the decoy state model to be able to address such \textit{lossy unitary} circuits.

\begin{figure}[h]
	\includegraphics[scale=0.3]{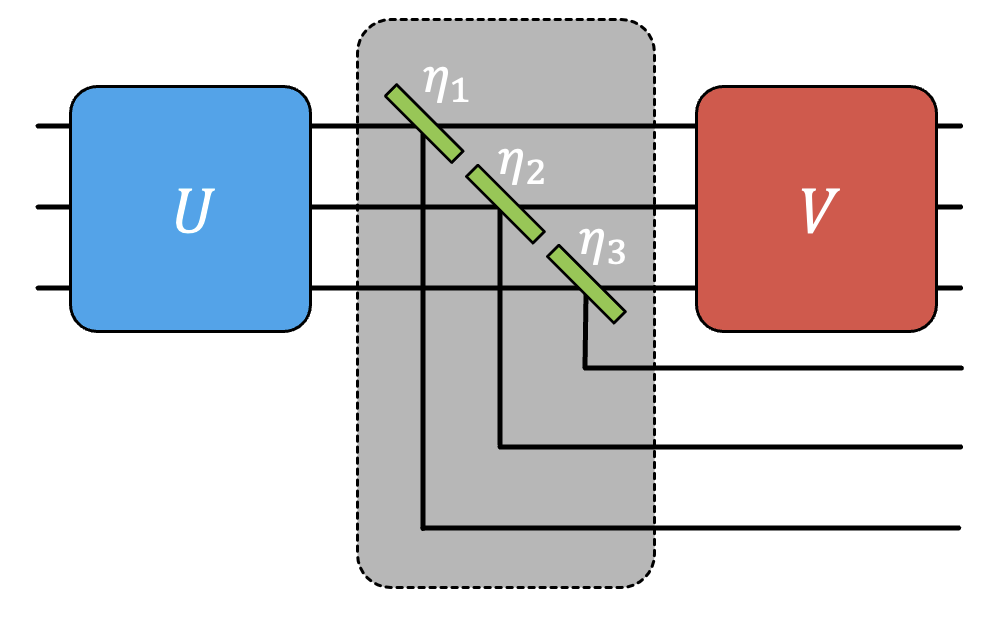}
	\caption{A general representation of a lossy $N\times N$ unitary circuit, which can always be expanded into $N$ couplers to environmental modes sandwiched by two unitary matrices $U$ and $V$. The overall transformation is a unitary matrix of $2N\times 2N$ (where we only observe the first $N$ input and $N$ output ports).}
	\label{fig:lossy_unitary}
\end{figure}

For this case, it is easy to simply replace the random unitary channels in the training data with random lossy unitary ones. As shown in Ref. \cite{lossy_unitary}, an $N \times N$ lossy unitary circuit can always be generally represented by a $2N\times 2N$ lossless unitary circuit (with $N$ additional environmental modes that are not observed), which consists of two unitaries and $N$ couplers between them, as shown in Fig. \ref{fig:lossy_unitary}. We therefore simply draw Haar-random $U$ and $V$ and allow the couplers to be random numbers in $[0,1]$. Note that, as this is used for training data of the linear regression, we do not need to prove that it uniformly draws from the entire possible space of random lossy unitaries, but just that it spans the space.

Note that this system still has a number of degrees-of-freedom that scales with $N^2$ (unlike the arbitrary channel scenario, where degrees-of-freedom depend on the huge photon number space), which means that here, too, the higher-order yields are locked to the lower-order yields with the physical prior of the system being ``lossy unitary".

\begin{figure}[h]
		\includegraphics[scale=0.6]{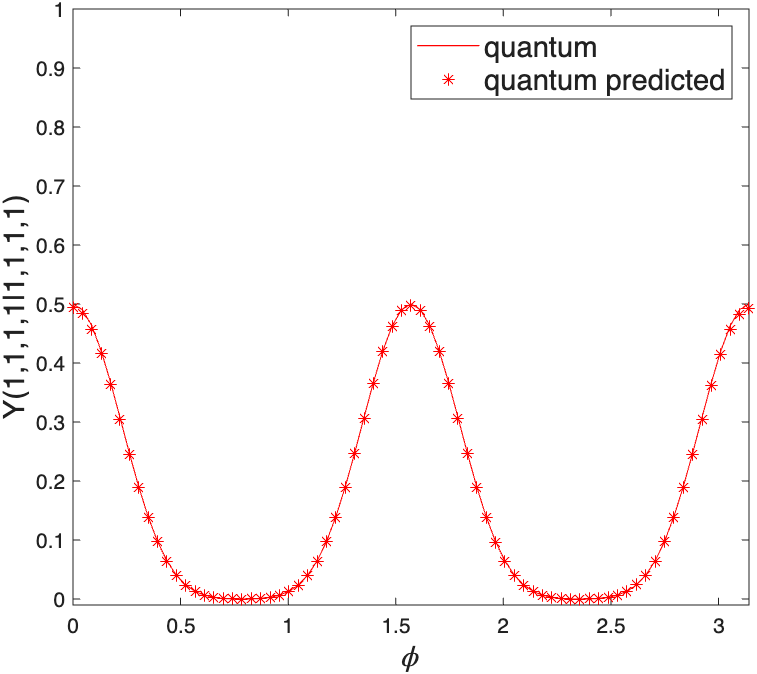}
		\caption{Estimation of $Y(1,1,1,1|1,1,1,1)$ for the same circuit as in Fig. \ref{fig:metrology}, except with losses of $1-\sqrt[4]{0.5}$ coupled to each input port (such that the overall yields are scaled by $0.5$). The linear regression model is trained with random lossy unitaries, and we use 64 LHS settings in this case. The resulting average absolute error of predictions is only about $8.8\times 10^{-5}$.}
		\label{fig:rate_lossy}
	\end{figure}

For instance, in Fig. \ref{fig:rate_lossy} we plot the same $4\times 4$ setup as Fig. \ref{fig:metrology} but with additional losses coupled to each input port, which can physically come from, for instance, insertion loss. As can be seen, the linear regression model trained on lossy unitary data can accurately predict $Y(1,1,1,1|1,1,1,1)$ as usual.

\subsection{Statistical Fluctuations}

Next, we consider statistical fluctuations, since we can only perform experiments in finite time. Let us assume we send a total of $L$ signals. For each particular decoy setting of $\vec{\mu},\vec{t}$, we denote $L_{\vec{\mu},\vec{t}}$ as the number of signals sent in this particular setting. The outcome $\vec{p}$ is not a user-chosen setting but is an event pattern, for which the gain for all combinations of $\vec{p}$ adds up to one, i.e.

\begin{equation}
	\sum_{\vec{p}} Q_{\vec{\mu},\vec{t},\vec{p}} = 1.
\end{equation}

In this case, the set of outcomes of $\vec{p}$ follow a multinomial distribution, which, in the limit of large numbers of signals sent, can be approximated by a multivariate Gaussian distribution. In simulation, we simply sample counts for each pattern from a 1-D normal distribution with expected value $L_{\vec{\mu},\vec{t}} \,Q_{\vec{\mu},\vec{t},\vec{p}}$ and standard deviation $(L_{\vec{\mu},\vec{t}} \, Q_{\vec{\mu},\vec{t},\vec{p}})^{1/2}$. If the total probabilities do not normalize, i.e. $\sum_{\vec{p}} Q_{\vec{\mu},\vec{t},\vec{p}} = 1 + \Delta$, the correction counts $-L_{\vec{\mu},\vec{t}} \, \Delta$ is distributed to all patterns $\vec{p}$ in this set of $\vec{\mu},\vec{t}$ depending on the weight (expected value) $L_{\vec{\mu},\vec{t}} \, Q_{\vec{\mu},\vec{t},\vec{p}}$.

For simplicity, we can set $L_{\vec{\mu},\vec{t}} = L / |\{\vec{\mu},\vec{t}\}|$ (where we define $|\{\vec{\mu},\vec{t}\}|$ as the total number of input/output decoy settings), which represents a uniform distribution of signals to all settings. In principle one can further optimize this by allocating more signals to settings with weaker input intensities and higher detector loss.

As one decreases the data size, there will be more statistical fluctuations in the observables $\{Q_{\vec{\mu},\vec{t},\vec{p}}\}$. We include such noise in both the training data and the testing data. Naturally, the more statistical noise present in the data, the less accurate the linear regression and the output prediction values will be. Nonetheless, in the unitary scenario we again remark that, due to the physical prior, the higher-order yields (such as $Y(1,1,1,1|1,1,1,1)$) are locked to lower-order ones, due to a unitary system having a limited number of degrees-of-freedom. This makes the higher-order yields highly resilient against noises compared with linear programming approach, Vandermonde inversion, or linear regression on arbitrary channels, as we will discuss next. 

\begin{figure}[h]
		\includegraphics[scale=0.6]{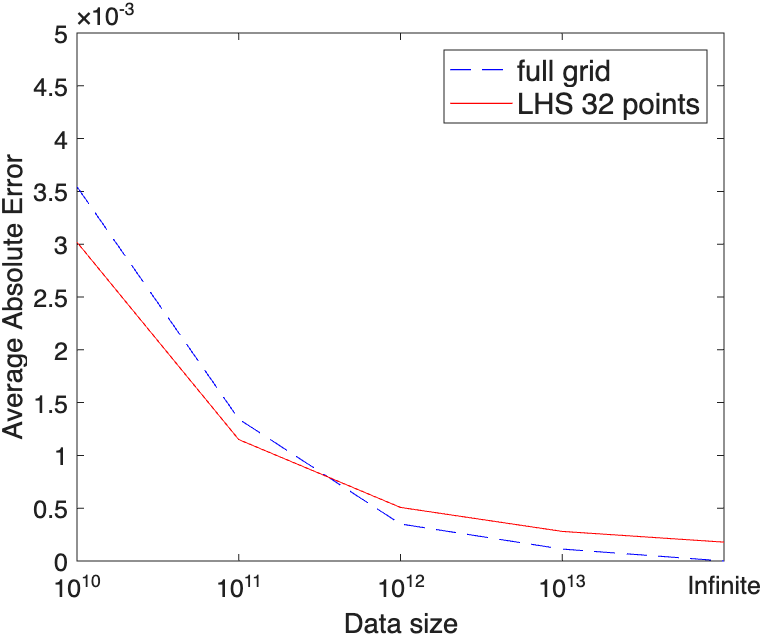}
		\caption{Average absolute error plotted against total sent data size for the same circuit as in Fig. \ref{fig:metrology}, estimating $Y(1,1,1,1|1,1,1,1)$. We assume equal distribution of data among all decoy settings. As can be seen, the model can still accurately predict the photon statistics even at data sizes of $10^{12}$ or $10^{11}$ (with absolute error on the order of $10^{-3}$). Also, note that as the data size gets smaller, using LHS starts to gain an advantage over using the full Cartesian grid, as the former uses much fewer settings and thus allows less statistical fluctuations due to each setting getting more data.}
		\label{fig:finite}
\end{figure}

As shown in Fig. \ref{fig:finite}, again using the $4\times 4$ circuit as an example, in the asymptotic case, using full Cartesian grid for $\vec{\mu},\vec{t}$ will enable near machine-level precision ($10^{-15}$ absolute error), while using 32-setting LHS yields about $1.8\times 10^{-4}$ absolute error. As data size decreases and more statistical noise is included, we see the absolute errors increase, but even at data sizes of $10^{12}$ or $10^{11}$ (which is experimentally very feasible – for instance a GHz system can generate $10^{12}$ signals in 17 minutes and $10^{13}$ in less than 3 hours), the absolute error is still maintained on the order of $10^{-3}$ or below. For a $6\times 6$ circuit (such as the setup for Fig. \ref{fig:metrology2}, using LHS settings), we need more data but can still get absolute errors of $1.7 \times 10^{-3}$ and $8.2 \times 10^{-4}$ at $L=10^{13}$ and $L=10^{14}$, respectively.

\begin{figure}[h]
		\includegraphics[scale=0.6]{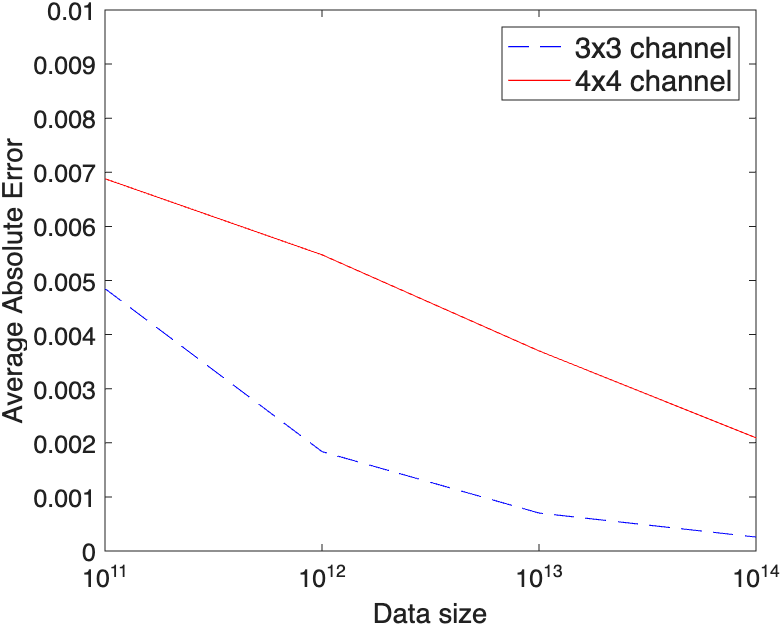}
		\caption{Average absolute error plotted against total sent data size for passive randomized 3-input-3-output and 4-input-4-output quantum channels, estimating $Y(1,1,1|1,1,1)$ and $Y(1,1,1,1|1,1,1,1)$. Here we still use the full Cartesian grid, with 4 input decoy settings $\{0,0.05,0.3,0.7\}$ and 2 output decoy settings $\{0.1,0.5\}$ for each port. As can be seen, the accuracy of the model is much more significantly impacted by statistical noise here, due to not using any physical prior and the inherent ill-conditionedness of the problem. Nonetheless, the linear regression model can still accurately predict the photon statistics for 3-input-3-output channels (with absolute error of $2\times 10^{-3}$) for a data size of $10^{12}$ or higher. For 4-input-4-output channels, the model is inherently unstable, but the linear regression still converges and predicts with an average absolute error of $3.7 \times 10^{-3}$ and $2.1\times 10^{-3}$ at data sizes of $10^{13}$ and $10^{14}$. Comparatively, in the asymptotic scenario, the average absolute errors for predicted $Y(1,1,1|1,1,1)$ and $Y(1,1,1,1|1,1,1,1)$ of random passive channels are $9.9\times 10^{-6}$ and $2.7\times 10^{-5}$.}
		\label{fig:finite_random}
\end{figure}

In contrast, statistical fluctuations will have a much higher impact on the models without physical prior. This is mainly because, as we mentioned in Sec. II.B, fundamentally the large $\left|{\partial Y(\vec{m}|\vec{n})}\over{\partial Q_i}\right|$ results in a very unstable mapping, where small changes in the observables $Q_i$ will have a big impact on predicted $Y$, even if we can obtain the inverse of coefficient matrix $A$ down to machine-precision or perfectly solve the linear program.  Nonetheless, we show that with our linear regression method, even in the presence of statistical noise and no channel assumption, we can accurately predict $Y(1,1,1|1,1,1)$ or predict $Y(1,1,1,1|1,1,1,1)$ with noise, as shown in Fig. \ref{fig:finite_random}. The 4-input-4-output case is inherently noisy due to the ill-conditioned nature of the problem (a fundamental limitation for the arbitrary channel scenario), but one can in principle further improve the results with an optimized LHS choice (since the full Cartesian grid improves accuracy but aggravates finite-size penalty) or with optimally allocated data sizes $L_{\vec{\mu},\vec{t}}$ (with more data sent under weaker-intensity settings), which will be the subject of future studies.

\subsection{Intensity Fluctuations}

Lastly, intensity fluctuation from laser sources or imperfect modulators, like observed for QKD setups \cite{intensity_fluctuation}, will result in an imperfect decoy state analysis, as the method always assumes that we have perfect knowledge of our decoy intensities in order to build the linear constraints. Note that here we are not working with an adversarial setup, so we do not need to prove the worst-case bounds in the presence of intensity fluctuation, nor do we have to worry about intensity correlations between neighboring pulses. Here, we are only interested in how much using observables obtained from fluctuating light sources will affect the accuracy of yield prediction. 

(1) Let us first consider fast, per-pulse fluctuations that change randomly from pulse to pulse (we assume they follow a normal distribution), which result in an integration of $\vec{\mu}$ over a normal distribution when generating each $Q_{\vec{\mu},\vec{t},\vec{p}}$. We find that, in such a scenario, the performance of the linear regression model is hardly affected (so long as the training data also includes the same fluctuations). For instance, for $Y(1,1,1,1|1,1,1,1)$ in random unitary channels, an ideal linear regression model (with a full Cartesian grid of $\{0.5,0.1\}$ input intensities and $\{0.5,0.1\}$ attenuations) can yield near machine precision with an average absolute error of $1.9\times 10^{-15}$. When a per-pulse intensity fluctuation is introduced, even with a standard deviation as large as $5\%$ of the expected intensity values, the average absolute error only increases to $1.3\times 10^{-14}$, which is still near machine precision if no other noise is present in the data other than intensity fluctuations. This is mainly because, as shown in Table II, the observables $Q_{\vec{\mu},\vec{t},\vec{p}}$ coming from phase-randomized coherent lights are naturally insensitive to symmetric fluctuations in intensities, so the sets of observables almost remain identical regardless of whether per-pulse fluctuations are present.

(2) Note that the above scenario is different from a slow, per-setting fluctuation, which results in slightly different deviations (one-time samples from a normal distribution) for each $\vec{\mu}$ when iterating over all input decoy settings $\{\vec{\mu}\}$. In this case, the deviated values are assumed to be unchanged throughout the data collection of this particular setting, essentially creating a bias for each $Q_{\vec{\mu},\vec{t},\vec{p}}$ in different directions. Unlike a per-pulse fluctuation (which causes symmetric noises), such a model will have a severe impact on the performance of the decoy state method. This is because the key assumption of decoy-state analysis is that the channel (i.e. the $Y(\vec{m}|\vec{n})$ to be solved) remains identical across decoy settings, but a per-setting deviation is equivalent to an attenuation in the channel that changes with the setting, which would result in contradictory constraint rows in the linear equations and would, in the worst case, cause a linear program to be infeasible. While the linear regression model would not directly fail due to infeasibility, its accuracy will be considerably affected. For instance, using the same example of $Y(1,1,1,1|1,1,1,1)$ in random unitary channels above, introducing just $1\%$ or $2\%$ of per-setting fluctuation would increase the absolute error to $2.0 \times 10^{-4}$ and $4.0 \times 10^{-4}$, respectively (although the model is still very usable at these levels of error). In practice, one can avoid such type of per-setting fluctuation by simply randomizing the decoy settings pulse-by-pulse, similar to what is done in QKD systems, but here it is not for security but to distribute intensity drifts evenly across settings.

\begin{table}[htbp]
	\label{table:intensity_fluctuation}
	\centering
	\caption{Effect of per-pulse intensity fluctuations on observable. We pick the circuit in Fig. \ref{fig:metrology} and choose fixed expected values of input intensities $\vec{\mu}=\{0.5,0.5,0.5,0.5\}$, fixed detector attenuations $\vec{t}=\{0.5,0.5,0.5,0.5\}$, and all-click events $\vec{p}=\{1,1,1,1\}$. We consider two scenarios of $\phi=\pi/2$ (peak) and $\phi=\pi/4$ (dip) and calculate the gain $Q$ from phase-randomized input coherent states against various fluctuation levels. Fluctuations are modeled as intensity integrations over a normal distribution with a given percentage of the expected value as the standard deviation. As can be seen, within reasonable fluctuation (such as up to $5\%$), the effect on observed gain is negligible (less than $0.5\%$).}
	\begin{tabular}{ccc}
		Fluctuation & Gain at $\phi=\pi/2$ & Gain at $\phi=\pi/4$\\
		\hline\hline
		None & $2.39406 \times 10^{-3}$ & $6.28721 \times 10^{-4}$ \\
		$1\%$ & $2.39395 \times 10^{-3}$ & $6.28808 \times 10^{-4}$\\
		$2\%$ & $2.39363 \times 10^{-3}$ & $6.29069 \times 10^{-4}$\\
		$3\%$ & $2.39311 \times 10^{-3}$ & $6.29503 \times 10^{-4}$\\
		$5\%$ & $2.39142 \times 10^{-3}$ & $6.30893 \times 10^{-4}$\\
		$10\%$ & $2.38354\times 10^{-3}$ & $6.37409 \times 10^{-4}$\\
	\end{tabular}
\end{table}

\subsection{Conclusion}

As we can see from above, our linear regression model is fairly robust against all three types of imperfections/noises (optical loss, statistical fluctuations, intensity fluctuations). In particular, for the unitary circuit scenario, optical loss and per-pulse intensity fluctuation have little impact on the prediction accuracy. Statistical fluctuation is probably the biggest source of error, but unitary models still work well under realistic data size constraints (such as $4\times 4$ with $10^{11}-10^{12}$ signals and even up to $6\times 6$ circuit with $10^{13}-10^{14}$ signals), while the model with arbitrary channel, as expected, is unstable at $4\times 4$ sizes if there is statistical fluctuation, due to the inherent ill-conditionedness of the problem. Nonetheless, it can still be used to accurately probe for instance nonlinear media while simulating three photons, which, again, is something better than prior art linear programming approaches.

\section{Training-Free Approach with Vandermonde Matrix Inversion}

	In this section, we propose an alternative training-free approach for the full input/output decoy analysis problem, which is built on the analytical inversion of Vandermonde matrix \cite{Xiongfeng_interference,Chau_Vandermonde,Hayashi_Vandermonde} combined with a least-square numerical optimization. We also compare it with the linear regression approach in the main text.

	We observe that the decoy state model in main text Sec. II.A, taking Eq. \eqref{eq:decoy_full} as an example, can be broken up into a two-step procedure of (1) solving for the statistics from specific input Fock states and (2) solving for the final target statistics with fixed input and output photons. The first step in analyzing the input states can be further unwrapped into a series of Vandermonde matrix inversions and analytically solved \cite{Xiongfeng_interference}, while we observe that the second step for output states has much smaller dimensions, so we can in fact easily solve it with a least-square numerical optimization.

	First, let us take a more careful look back at Eq. \eqref{eq:decoy_full}:
	
	\begin{equation*}\label{eq:decoy_full_2}
		Q_{\vec{\mu},\vec{\eta},\vec{p}} = \sum_{\substack{\vec{n},\vec{m}}} P_{\vec{\mu}}(\vec{n}) D_{\vec{\eta},\vec{p}}(\vec{m}) Y(\vec{m}|\vec{n}).
	\end{equation*}
	
	\noindent using which, suppose one does an experiment with a full set of input decoy and output decoy settings $\{\vec{\mu},\vec{\eta},\vec{p}\}$, one can in principle solve any target yield $Y(\vec{m}_{\text{target}}|\vec{n}_{\text{target}})$ given enough number of decoy settings, in the ideal case where there are no numerical and experimental noises.
	
	Now, we can rewrite the equation as:
	
		\begin{equation}\label{eq:first_decoy}
			\begin{aligned}
					Q_{\vec{\mu},\vec{\eta},\vec{p}} &= \sum_{\vec{n}} P_{\vec{\mu}}(\vec{n}) \left [  \sum_{\vec{m}} D_{\vec{\eta},\vec{p}}(\vec{m}) Y(\vec{m}|\vec{n}) \right ]\\
					&= \sum_{\vec{n}} P_{\vec{\mu}}(\vec{n})Y_{\vec{\eta},\vec{p}}(\vec{n}),
			\end{aligned}
	\end{equation}
	
	\noindent which means that the user first \textit{groups} the data with the same $\{\vec{\eta},\vec{p}\}$ settings and solves for specific input-only yields for each specific $\vec{\eta},\vec{p}$, in the same fashion as Sec. II.A point 2. In this step 1, the user will have to solve the linear program multiple times (once for each $\vec{\eta},\vec{p}$ in the total list of settings), obtaining a set of $Y_{\vec{\eta},\vec{p}}(\vec{n}_{\text{target}})$ for each $\vec{\eta},\vec{p}$. Then, in step 2, they can perform the decoy state analysis a second time using
	
	\begin{equation}\label{eq:second_decoy}
		\begin{aligned}
			Y_{\vec{\eta},\vec{p}}(\vec{n}_{\text{target}}) &= \sum_{\vec{m}}  D_{\vec{\eta},\vec{p}}(\vec{m}) Y(\vec{m}|\vec{n}_{\text{target}}),
		\end{aligned}
	\end{equation}

	\noindent which combines the various $\vec{\eta},\vec{p}$ settings as constraints to obtain the specific $\vec{m}_{\text{target}}$ term, i.e. the final output $Y(\vec{m}_{\text{target}}|\vec{n}_{\text{target}})$. Note that in this linear program, $\vec{n}_{\text{target}}$ is a known number that is already fixed in step 1, and the left-hand-side terms $Y_{\vec{\eta},\vec{p}}(\vec{n}_{\text{target}})$ obtained from step 1 are treated as the new set of observables (i.e. equivalent to the observables we would have gotten had we used Fock state sources).
	
	Furthermore, one can also separate each element in $\vec{n}_{\text{target}}$ in step 1 and perform a 1-D decoy analysis one-at-a-time using statistics from various input decoy settings $\mu_i$ at each port, while keeping the other intensities and $\vec{\eta},\vec{p}$ frozen. Using the same notations as Sec. II.A point 3, we can write
	
	\begin{equation}\label{eq:decoy_input_partial}
		Q_{\vec{\mu},\vec{\eta},\vec{p}} = \sum_{\substack{\vec{n_I}}} P_{\vec{\mu}}(\vec{n_I}) Y_{\vec{\mu}_{\overline{I}},\vec{\eta},\vec{p}}(\vec{n_I}),
	\end{equation}
	
	\noindent which means that the input decoy analysis can be performed on arbitrary ports to only track the photon numbers in those selected points in set $I$, while treating the intensities at untracked ports (in the complement set $\overline{I}$ in the set of all indices $[N]$) as constants. If we set $I$ to be just port 1, we have

	\begin{equation}\label{eq:single_decoy}
		\begin{aligned}
			Q_{\vec{\mu},\vec{\eta},\vec{p}} &= \sum_{n_1} P_{\mu_1}(n_1)Y_{\vec{\mu}_{\backslash 1},\vec{\eta},\vec{p}}(n_1),
		\end{aligned}
	\end{equation}
	
	\noindent where we use a simplified notation $\vec{\mu}_{\backslash 1}$ to denote $\vec{\mu}_{\overline{\{1\}}}$, i.e. the vector $\vec{\mu}$ excluding the $\mu_1$ element, i.e. $\{\mu_2,\mu_3,...\}$. The above equation solves for the term with $n_1$ by combining the data from various $\mu_1$ settings, and this process is repeated for every setting of $\{\vec{\mu}_{\backslash 1},\vec{\eta},\vec{p}\}$ (each one of which is a linear program). We can then use the set of obtained $\{{Y_{\vec{\mu}_{\backslash 1},\vec{\eta},\vec{p}}(n_{1,\text{target}})}\}$ as the new left-hand-side ``observables" and iterate over say $\{\mu_2\}$ to get $\{{Y_{\vec{\mu}_{\backslash 1,2},\vec{\eta},\vec{p}}(n_{1,\text{target}},n_{2,\text{target}})}\}$:
	
	\begin{equation}\label{eq:single_decoy_2}
		\begin{aligned}
			Y_{\vec{\mu}_{\backslash 1},\vec{\eta},\vec{p}}(n_1) &= \sum_{n_2} P_{\mu_2}(n_2)Y_{\vec{\mu}_{\backslash 1,2},\vec{\eta},\vec{p}}(n_{1,\text{target}},n_2),
		\end{aligned}
	\end{equation}
	
	\noindent where $n_{1,\text{target}}$ is a constant. Afterwards, we can then further iterate over $\{\mu_3\}$ to obtain $\{{Y_{\vec{\mu}_{\backslash 1,2,3},\vec{\eta},\vec{p}}(n_{1,\text{target}},n_{2,\text{target}},n_{3,\text{target}})}\}$, and so on.\\
	
	For each of the 1-D decoy state analysis (taking Eq. \eqref{eq:single_decoy} as an example), note that the Poisson coefficients $P_{\mu_1}(n_1)=e^{-\mu_1}{{\mu_1^n}\over{n_1!}}$ can be re-ordered into just a power polynomial for $\mu_1$:
	
	\begin{equation}\label{eq:interpolation}
		\begin{aligned}
			e^{\mu_1} Q_{\vec{\mu},\vec{\eta},\vec{p}} &= \sum_{n_1} \left[ {Y_{\vec{\mu}_{\backslash 1},\vec{\eta},\vec{p}}(n_1)} \over {n_1 !}\right] \left (\mu_1^{n_1} \right ).
		\end{aligned}
	\end{equation}
	
	If we write this out into a matrix $A_1$ whose rows correspond to $\mu_1$ and columns correspond to $n_1$, we can see it is a Vandermonde matrix \cite{Chau_Vandermonde,Xiongfeng_interference,Hayashi_Vandermonde}, which exhibits a very nice property of having an explicit analytical inverse, which means that the linear program Eq. \eqref{eq:interpolation} can be solved analytically by simply inverting the matrix $A_1$. For each element in $\vec{\mu}$, one can solve a series of Vandermonde matrix inversions one-at-a-time, eventually obtaining $Y_{\vec{\eta},\vec{p}}(\vec{n}_{\text{target}})$.
	
	To obtain the final output $Y(\vec{m}_{\text{target}}|\vec{n}_{\text{target}})$, we solve a single set of linear equations Eq. \eqref{eq:second_decoy} using the set of $\{\vec{\eta},\vec{p}\}$ and the corresponding $Y_{\vec{\eta},\vec{p}}(\vec{n}_{\text{target}})$ for each output decoy setting we obtained from step 1. Note that here the dimensionality is much smaller, since photon numbers $\sum m_i$ has to be equal to (or in the case of lossy channels, less than or equal to) $\sum n_i$, and some or all of the coefficients $D_{\vec{\eta},\vec{p}}(\vec{m})$ where $p_i=1$ also do not carry the exponentially vanishing tail like $P_{\vec{\mu}}(\vec{n})$ (but rather approach one as photon numbers increase).
	
	Nonetheless, using a linear programming solver is still not going to be effective here, since the numerical errors (both numerically from the inherent numerical inaccuracy in inverting Vandermonde matrix with large coefficients, and experimentally from measurement inaccuracy and statistical fluctuations) accumulate each time we apply a Vandermonde matrix inversion, and such errors will most of the time render the linear program infeasible. Here we propose a simple method for solving Eq. \eqref{eq:second_decoy} in the presence of numerical noise: considering that we only need to \textit{estimate} the target yield, instead of obtaining provable upper/lower bounds, we can simply perform a least-square optimization instead of a linear program to find the best approximate solution. More specifically, we optimize the vector $\vec{Y}_m = \vec{Y}(\vec{m}|\vec{n}_{\text{target}})$ consisting of yields for all eligible $\vec{m}$, subject to the constraint that $\sum_{\vec{m}} Y(\vec{m}|\vec{n}_{\text{target}}) = 1$, such that the squared-distance between $D \vec{Y}$ and $\vec{Y}_{\vec{\eta},\vec{p}}$ is the smallest, where $D$ is the matrix whose elements are $D_{\vec{\eta},\vec{p}}(\vec{m})$, and $\vec{Y}_{\vec{\eta},\vec{p}}(\vec{n}_{\text{target}})$ is the step 1 yield results (treated as the new observables) whose rows correspond to settings of $\{\vec{\eta},\vec{p}\}$.
	
	\begin{equation}
		\begin{aligned}
			\text{minimize} \quad& ||\vec{Y}_{\vec{\eta},\vec{p}}-D \vec{Y}_m||^2,\\
			\text{subject to}  \quad & \sum_{\vec{m}} Y(\vec{m}|\vec{n}_{\text{target}})  =1, \\
			& 0\leq Y(\vec{m}|\vec{n}_{\text{target}}) \leq 1.\\
		\end{aligned}
	\end{equation}
	
	\noindent In practice this can be performed conveniently with tools such as the Matlab built-in function ``lsqlin". The key point is that, such an optimization does not force the solution to perfectly satisfy all constraints, but tries to find the closest solution, while a linear programming solver will report infeasibility and give up. Given the considerable amount of numerical noise, the optimization gives us a close enough solution for practical use.

		\begin{figure}[h]
		\includegraphics[scale=0.6]{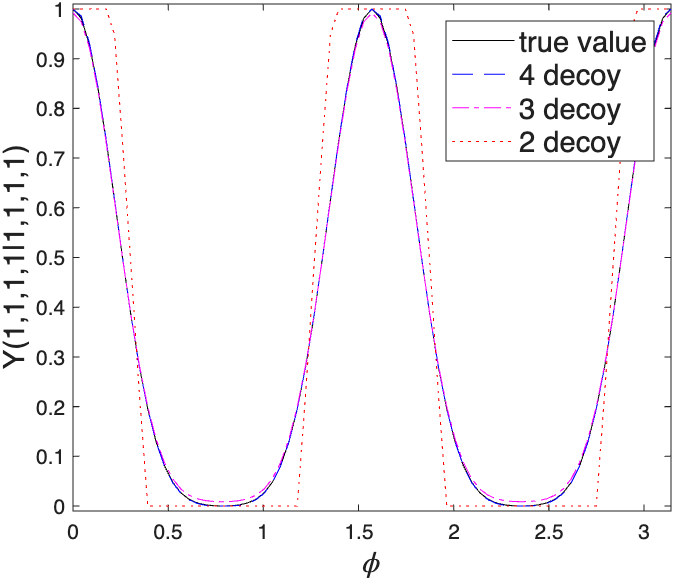}
		\caption{Estimation of $Y(1,1,1,1|1,1,1,1)$ for the same circuit as in Fig. \ref{fig:metrology}, with the Vandermonde+optimization two-step method. Here we use a full Cartesian grid (as required by the two-step method) including two output attenuator settings of $\{0.5,0.1\}$ for each port and a range of $2$,$3$, or $4$ input decoy settings $\{0.1,0.5\}$, $\{0,0.05,0.3\}$, $\{0,0.05,0.3,0.7\}$ for each port. As can be seen here, the two-step approach needs as many as 4 input decoy settings to be comparable with results in Fig. \ref{fig:metrology}. The average absolute error tested over Haar-random unitaries is $2.8 \times 10^{-2}$ for 2 decoys, $7.8 \times 10^{-3}$ for 3 decoys, and $8.5 \times 10^{-4}$ for 4 decoys (all of which mean the number of input decoy settings per port).}
		\label{fig:rate_Vandermonde}
	\end{figure}
	
	Note that, while the Vandermonde matrix inversion is much faster and more accurate/stable than the traditional linear program approach, it does not magically solve the problem mentioned in Sec. II.B that estimated yields are highly sensitive to noises in the observed gains, i.e. $\left|{\partial Y(\vec{m}|\vec{n})}\over{\partial Q_i}\right|$ is large. Such a problem is manifested here as the numerical instability of Vandermonde matrix itself, which has very large coefficients. As this approach does not make any assumptions (such as unitary assumption) on the channel, it will be heavily limited by the dimensionality of the problem and will require more decoy settings to get an accurate result.

	For instance, in Fig. \ref{fig:rate_Vandermonde}, we can see that, with simulated noiseless observables, to reproduce similar prediction accuracy with the two-step training-free approach, we need to use up to 4 input decoy settings, such as $\{0,0.05,0.3,0.7\}$, but the linear regression approach (which uses a physical prior) can achieve near machine precision with just 2 input decoy settings $\{0.5,0.1\}$. On the other hand, even for random passive channels, the two-step approach still yields worse accuracy than a linear regression model trained on random channel data when using the same number of decoy settings, as shown in Table III. Therefore, in the main text, we picked the linear regression model as the main approach we would like to advocate.

	\begin{table}[htbp]
		\label{table:random_Vandermonde}
		\centering
		\caption{Examples of predicted data from linear regression versus two-step method (and versus true yields) for randomized passive 4-input-4-output channels (top), which are the same testing channels as in Table I, and 3-input-3-output channels (bottom). The full Cartesian product grid with 4 input intensities $\{0,0.05,0.3,0.7\}$ is used here. Data is presented with 5 decimal places and the absolute error is rounded to 2 significant digits. The average absolute errors for the linear regression and two-step models (tested over 100 random passive channels) are respectively $2.7\times 10^{-5}$ and $1.0\times 10^{-3}$ for the $4\times4$ case as well as $9.9\times 10^{-6}$ and $5.2\times 10^{-4}$ for the $3\times3$ case.}
		\begin{tabular}{ccccc}
			$Y_{\text{true}}^{4\times 4}$ & $Y_{\text{regression}}^{4\times 4}$ & Abs. Err. & $Y_{\text{two-step}}^{4\times 4}$ & Abs. Err.\\
			\hline
			$0.01918$ & $0.01919$ & $5.2\times 10^{-6}$ & $0.01809$ & $1.1\times 10^{-3}$\\
			$0.00212$ & $0.00215$ & $3.0\times 10^{-5}$ & $0.00132$ & $8.0\times 10^{-4}$\\
			$0.00510$ & $0.00511$ & $1.7\times 10^{-6}$ & $0.00403$ & $1.0\times 10^{-3}$\\
			$0.02200$ & $0.02202$ & $2.9\times 10^{-5}$ & $0.02087$ & $1.1\times 10^{-3}$\\
			$0.02746$ & $0.02743$ & $3.2\times 10^{-5}$ & $0.02630$ & $1.2\times 10^{-3}$\\
			\\
			$Y_{\text{true}}^{3\times 3}$ & $Y_{\text{regression}}^{3\times 3}$ & Abs. Err. & $Y_{\text{two-step}}^{3\times 3}$ & Abs. Err.\\
			\hline
			$0.06882$ & $0.06880$ & $2.0\times 10^{-5}$ & $0.06834$ & $4.8\times 10^{-4}$\\
			$0.00057$ & $0.00058$ & $1.2\times 10^{-5}$ & $0.00037$ & $2.0\times 10^{-4}$\\
			$0.08402$ & $0.08401$ & $2.9\times 10^{-6}$ & $0.08350$ & $5.2\times 10^{-4}$\\
			$0.01184$ & $0.01186$ & $2.3\times 10^{-5}$ & $0.01134$ & $5.0\times 10^{-4}$\\
			$0.00488$ & $0.00487$ & $9.4\times 10^{-6}$ & $0.00427$ & $6.1\times 10^{-4}$\\
		\end{tabular}
	\end{table}

\newpage
	

\begin{thebibliography}{}

	\bibitem{KLM} E Knill, R Laflamme, GJ Milburn. ``A scheme for efficient quantum computation with linear optics." nature 409.6816 (2001): 46-52.


	\bibitem{onewaycomputing} R Raussendorf, HJ Briegel. ``A one-way quantum computer." Physical review letters 86.22 (2001): 5188.

	\bibitem{bosonsampling1} H Wang, \emph{et al.} ``High-efficiency multiphoton boson sampling." Nature Photonics 11.6 (2017): 361-365.
	\bibitem{bosonsampling2} HS Zhong, \emph{et al.} ``Quantum computational advantage using photons." Science 370.6523 (2020): 1460-1463.

	\bibitem{bell1} A Aspect, J Dalibard, G Roger. ``Experimental test of Bell's inequalities using time-varying analyzers." Physical review letters 49.25 (1982): 1804.
	
	\bibitem{bell2} G Weihs, \emph{et al.} ``Violation of Bell's inequality under strict Einstein locality conditions." Physical Review Letters 81.23 (1998): 5039.

	\bibitem{bell3} LK Shalm, \emph{et al.} ``Strong loophole-free test of local realism." Physical review letters 115.25 (2015): 250402.

	\bibitem{metrology} V Giovannetti, S Lloyd, L Maccone. ``Advances in quantum metrology." Nature photonics 5.4 (2011): 222-229.
	
	\bibitem{quantumdot} DA Vajner, \emph{et al.} ``Quantum communication using semiconductor quantum dots." Advanced Quantum Technologies 5.7 (2022): 2100116.
	
	\bibitem{diamondcenter} I Aharonovich, \emph{et al.} ``Diamond-based single-photon emitters." Reports on progress in Physics 74.7 (2011): 076501.

	\bibitem{Xiongfeng_interference} X Yuan, Z Zhang, N Lütkenhaus, X Ma, ``Simulating single photons with realistic photon sources." Physical Review A, 94, 062305 (2016).
	
	\bibitem{Marcos_interference} A Navarrete, W Wang, F Xu, M Curty, ``Characterizing multi-photon quantum interference with practical light sources and threshold single-photon detectors." New Journal of Physics 20.4 (2018): 043018.
	

	\bibitem{decoystate1} WY Hwang, ``Quantum key distribution with high loss: toward global secure communication." Physical Review Letters 91.5 (2003): 057901.
	\bibitem{decoystate2} HK Lo, XF Ma, and K Chen, ``Decoy state quantum key distribution." Physical review letters 94.23 (2005): 230504.
	\bibitem{decoystate3} XB Wang, ``Beating the photon-number-splitting attack in practical quantum cryptography." Physical review letters 94.23 (2005): 230503.

	\bibitem{this_companion} W Wang, HF Chau, ``Decoy-State Quantum Information Processing with Certified Bounds'', manuscript in preparation (2026).

	\bibitem{MDIQKD} HK Lo, M Curty, B Qi. ``Measurement-device-independent quantum key distribution." Physical Review Letters 108.13 (2012): 130503.
	
	\bibitem{CKA} Y Fu, et al. ``Long-distance measurement-device-independent multiparty quantum communication."  Physical Review Letters 114.9 (2015): 090501.

	\bibitem{LHS} MD McKay, RJ Beckman, WJ Conover. ``A comparison of three methods for selecting values of input variables in the analysis of output from a computer code." Technometrics 42.1 (2000): 55-61.



	
	\bibitem{Feihu_bosonsampling} F Xu, ``decoy state boson sampling", unpublished report (2014).

	\bibitem{ML1} W Wang, and HK Lo. ``Machine learning for optimal parameter prediction in quantum key distribution." Physical Review A 100.6 (2019): 062334.
	
	\bibitem{ML2} FY Lu, \emph{et al.} ``Parameter optimization and real-time calibration of a measurement-device-independent quantum key distribution network based on a back propagation artificial neural network." Journal of the Optical Society of America B 36.3 (2019): B92-B98.
	
	\bibitem{ML3} HJ Ding, \emph{et al.} ``Predicting optimal parameters with random forest for quantum key distribution." Quantum Information Processing 19 (2020): 1-8.
	
	\bibitem{ML4} MG Zhou, \emph{et al.} ``Neural network-based prediction of the secret-key rate of quantum key distribution." Scientific Reports 12.1 (2022): 8879.
	
	\bibitem{ML5} Q Dong, \emph{et al.} ``Optimization parameter prediction-based XGBoost of TF-QKD." Quantum Information Processing 21.7 (2022): 233.
	
	\bibitem{ML6} JL Kang, \emph{et al.} ``Machine learning with neural networks for parameter optimization in twin-field quantum key distribution." Quantum Information Processing 22.8 (2023): 309.
	
	\bibitem{ML7} L Wang, Q Dong, and R Jiao. ``Parameter optimization in decoy-state phase-matching quantum key distribution." Quantum Information Processing 22.10 (2023): 373.

	\bibitem{discrete_phase_randomization} Z Cao, Z Zhang, HK Lo, X Ma, ``Discrete-phase-randomized coherent state source and its application in quantum key distribution." New Journal of Physics 17, 053014 (2015).

	\bibitem{permanent1} S Scheel. ``Permanents in linear optical networks." arXiv preprint arXiv:0406127 (2004).

	\bibitem{permanent2} S Aaronson, A Arkhipov. ``The computational complexity of linear optics." Proceedings of the forty-third annual ACM symposium on Theory of computing. 2011.
	
	\bibitem{rounding_error} NJ Higham, ``Accuracy And Stability Of Numerical Algorithms", 2nd ed., SIAM Press (2002): 82.

	\bibitem{Chau_Vandermonde} HF Chau. ``Decoy-state quantum key distribution with more than three types of photon intensity pulses." Physical Review A 97.4 (2018): 040301.
	
	\bibitem{Hayashi_Vandermonde} M Hayashi, ``General theory for decoy-state quantum key distribution with an arbitrary number of intensities." New Journal of Physics 9.8 (2007): 284.

	\bibitem{metrology1} ZE Su, \emph{et al.} ``Multiphoton interference in quantum fourier transform circuits and applications to quantum metrology." Physical review letters 119.8 (2017): 080502.
	
	\bibitem{NOON} I Forbes, \emph{et al.} ``Heralded generation of entanglement with photons." arXiv preprint arXiv:2502.00982 (2025).

	\bibitem{lossy_unitary} N Tischler, C Rockstuhl, K Słowik. ``Quantum optical realization of arbitrary linear transformations allowing for loss and gain." Physical Review X 8, 021017 (2018).
	
	\bibitem{intensity_fluctuation} A Huang, A Mizutani, HK Lo, V Makarov, K Tamaki. ``Characterization of state-preparation uncertainty in quantum key distribution." Physical Review Applied 19, 014048 (2023).

	

	
	

	
	

\end{thebibliography}
\end{document}